\documentclass[aps,prd,twocolumn,titlepage,nofootinbib]{revtex4}
\usepackage{graphicx,physics}
\usepackage{color}
\usepackage{dcolumn}
\usepackage{latexsym}

\usepackage[normalem]{ulem}
\usepackage{hyperref,amssymb}
\usepackage{url}
\usepackage{color,soul}
\usepackage{graphicx}
\usepackage{verbatim}
\usepackage{multirow}
\usepackage{amsmath}
\newcommand{\beq}{\begin{eqnarray}}
\newcommand{\eeq}{\end{eqnarray}}
\usepackage{mathrsfs}
\usepackage{float,soul}
\usepackage[dvipsnames]{xcolor}
\usepackage{mathtools}
\usepackage{slashed}
\usepackage{physics}	
\usepackage{graphicx}   
\usepackage{epstopdf}
\usepackage{subfigure}  
\usepackage{hyperref}   
\usepackage{bbold}
\usepackage{wasysym}
\usepackage{feynmp}
\usepackage{hyperref}
\hypersetup{colorlinks,
}
\usepackage{tikz}

\def\ri{\rm i}

\usepackage[latin1]{inputenc}

\begin{document}

\title{Simple holographic dual of the Maxwell-Cattaneo model\\ \& the fate of KMS symmetry for non-hydrodynamic modes}
\author{Yongjun Ahn$^{1,2}$}
\author{Matteo Baggioli$^{1,2}$}
\email{b.matteo@sjtu.edu.cn}
\author{Yanyan Bu$^{3}$}
\email{yybu@hit.edu.cn}
\author{Masataka Matsumoto$^{4,1}$}
\author{Xiyang Sun$^{3}$}
\address{$^1$Wilczek Quantum Center, School of Physics and Astronomy, Shanghai Jiao Tong University, Shanghai 200240, China}
\address{$^2$Shanghai Research Center for Quantum Sciences, Shanghai 201315, China}
\address{$^3$School of Physics, Harbin Institute of Technology, Harbin 150001, China}
\address{$^4$Department of Physics, Chuo University, Tokyo 112-8551, Japan}

\begin{abstract}
Diffusion, as described by Fick's laws, governs the spreading of particles, information, data, and even financial fluctuations. However, due to its parabolic structure, the diffusion equation leads to an unphysical prediction: any localized disturbance instantaneously affects the entire system. The Maxwell-Cattaneo (MC) model, originally introduced to address relativistic heat conduction, refines the standard diffusion framework by incorporating a finite relaxation time $\tau$, associated with the onset of local equilibrium. This modification yields physically relevant consequences, including the emergence of propagating shear waves in liquids and second sound in solids. Holographic methods have historically provided powerful tools for describing the hydrodynamics of strongly correlated systems. However, they have so far failed to capture the dynamics governed by the MC model, limiting their ability to model intermediate time-scale phenomena. In this work, we construct a simple holographic dual of the Maxwell-Cattaneo model and rigorously establish its equivalence through a combination of analytical and numerical techniques. As an important byproduct of our analysis, and contrary to previous ad-hoc assumptions, we find that effective field theories featuring non-hydrodynamic modes exhibit a generalized form of Kubo-Martin-Schwinger (KMS) symmetry, which reduces to the canonical form only in the hydrodynamic limit.
\end{abstract}
\maketitle
\section{Introduction}
Diffusion describes the spreading of a continuous density field $\Phi(t, \vec{x})$ (\textit{e.g.}, ink concentration in a glass of water) that is governed by the following parabolic partial differential equation:
\begin{equation}
    \frac{\partial \Phi(t,\vec{x})}{\partial t}= D \nabla^2   \Phi(t,\vec{x}),\label{eq1}
\end{equation}
where $D$ is the diffusion coefficient \cite{10.1063/1.3177228}.
While physically diffusion originates from the random Brownian motion of the microscopic constituents, its universal character can be better appreciated from a macroscopic perspective where it can be derived by combining the \textit{continuity equation}, $\partial_t \Phi(t,\vec{x}) +\nabla \cdot \vec{j}(t,\vec{x})=0$, with the phenomenological Fick's law for the flux, $\vec{j}(t,\vec{x})=- D \nabla \Phi(t,\vec{x})$ \cite{Fick01071855}.

The diffusion equation \eqref{eq1} is widely used in physical, biological, geological, and social sciences. It is nevertheless well known, and first encountered in the so-called \textit{heat conduction paradox} \cite{RevModPhys.61.41,Jou_1988,10.1115/1.3143705}, that the diffusion equation does not correctly capture short-time dynamics, as it neglects the time lag required to establish a steady current $\vec{j}$ in a volume element once a gradient $\nabla \Phi$ has been imposed across it.

The Maxwell-Cattaneo (MC) model \cite{1573950400546202112,doi:10.1098/rstl.1867.0004} provides a simple solution to this problem \cite{1571980075311404160} by extending Fick's law to a more general form:
\begin{equation}
    \left(1+ \tau \partial_t\right) \vec{j}(t,\vec{x})=- D \nabla \Phi(t,\vec{x}),\label{eq1b}
\end{equation}
where $\tau$ is a relaxation time-scale characterizing the onset of local thermal equilibrium. Combined with the continuity equation, Eq.~\eqref{eq1b} yields to a hyperbolic partial differential equation:
\begin{equation}
   \tau  \frac{\partial^2 \Phi(t,\vec{x})}{\partial t^2}+ \frac{\partial \Phi(t,\vec{x})}{\partial t}= D \nabla^2   \Phi(t,\vec{x}),\label{eq2}
\end{equation}
that is known in the literature as the telegrapher \cite{heaviside1894electrical} or $k$-gap equation \cite{BAGGIOLI20201}, and that reduces to the diffusion equation \eqref{eq1} only for times $t \gg \tau$.

Eq.~\eqref{eq2} predicts the emergence of propagating excitations for times $t \ll \tau$ with speed $v_0^2=D/\tau$. This phenomenon is experimentally confirmed in several circumstances, including shear waves in liquid systems \cite{PhysRevLett.97.115001,Jiang2025,bai2025tracking}, second sound in solids \cite{PhysRev.131.2013}, and density fluctuations in cold atoms \cite{doi:10.1126/science.aat4134}.

Motivated by the experimental progresses in heavy-ion collisions \cite{Romatschke:2017ejr,Gale:2013da}, relativistic hydrodynamics \cite{Kovtun:2012rj} has emerged as a universal effective framework to capture late-time and long-wavelength dynamics in relativistic systems near thermal equilibrium. In this language, the continuity equation stems directly from the presence of a U(1) global symmetry, leading to the conservation of a four-current $J^\mu$ with $\mu=t,\vec{x}$. Importantly, diffusion arises also in neutral relativistic fluids where it describes the dynamics of shear fluctuations related to the stress tensor $T^{\mu\nu}$ and governed by the shear viscosity $\eta$ \cite{Kovtun:2012rj}. Also in this context, in order to retain causality, the diffusion equation has to be modified within the so-called Israel-Stewart (IS) formalism \cite{Israel:1979wp} which can somehow be thought as the tensorial version of the MC model.

Both the MC and IS frameworks can be successfully embedded in field theory language \cite{Jain:2023obu,Hongo:2024brb} (see also \cite{Baggioli:2020whu}) and are part of a larger program (sometimes labeled \textit{quasi-hydrodynamics} \cite{Martinoia:2024cbw}) aimed to extend hydrodynamics introducing non-hydrodynamic modes. Aside from the fundamental perspective, the introduction of non-hydro modes is fundamental near critical points \cite{PhysRevD.98.036006,PhysRevD.106.036017} and at shorter timescales \cite{PhysRevLett.130.212303}. 

The MC model was originally proposed to describe heat conduction in real materials. However, in most materials, the relaxation time $\tau$ is extremely short, typically on the order of picoseconds, so the non-hydrodynamic mode is often neglected~\cite{10.1115/1.3098984}. It is therefore instructive to identify physical scenarios in which a genuinely long-lived non-hydrodynamic mode is present. Although providing a general microscopic characterization of non-hydrodynamic modes remains a significant challenge, particularly in the presence of hydrodynamic fluctuations, two simple and illustrative examples can be discussed.

A first illustrative case is that of critical points \cite{Stephanov:2017ghc}. The simplest example involves a system with spontaneously broken U(1) symmetry (i.e., a superfluid) as the temperature approaches the critical value $T_c$. In this regime, the amplitude mode, associated with fluctuations in the magnitude of the order parameter, acquires a purely imaginary dispersion relation $\omega = - {\rm i} \Gamma$, with $\Gamma \propto (T_c - T)$. As $T \to T_c$, this non-hydrodynamic mode becomes increasingly long-lived and must be included in any accurate effective description (see, for instance, \cite{Donos:2022xfd,Donos:2022qao,Donos:2023ibv} in holographic and SK contexts). A similar scenario arises in QCD near its critical point, where extended hydrodynamic frameworks accounting for such long-lived non-hydro modes have been proposed~\cite{Stephanov:2017wlw,Stephanov:2017ghc}. Interestingly, within this framework, the microscopic origin of the non-hydrodynamic mode becomes clear: it corresponds to the Wigner transform of the equal-time two-point function of the fluctuations of the order parameter field.

A second ubiquitous and conceptually simple scenario involves systems with approximate symmetries or weakly broken conservation laws, as captured within the quasi-hydrodynamic framework~\cite{Grozdanov:2018fic}. In such cases, the coupling responsible for the explicit breaking of the symmetry is small compared to all other relevant scales in the system (e.g., the temperature), resulting in a nearly conserved quantity whose associated mode is long-lived and must be included in the effective theory.

A textbook example is electron transport in clean metals, where Umklapp scattering is weak and electron momentum decays very slowly, this is the essence of the Drude model. Another well-known example is the slow decay of the axial charge, which is governed by the smallness of the quark mass.

The AdS-CFT correspondence, or in short `\textit{holography}', has played and keeps playing a major role in the development of relativistic hydrodynamics \cite{GiuseppePolicastro_2002}, in particular when applied to strongly coupled systems such as the quark-gluon plasma \cite{Casalderrey-Solana:2011dxg}. While it is rather straightforward to derive \textit{vanilla} first-order relativistic hydrodynamics from holography, the study of quasi-hydrodynamics has emerged only more recently (\textit{e.g.}, \cite{Grozdanov:2018fic,Baggioli:2023tlc,Liu:2024tqe,Martinoia:2024cbw,Brattan:2024dfv}). In fact, despite these developments, a simple holographic model that fully captures the physical content and dynamics of the Maxwell-Cattaneo (MC) model is still lacking. To the best of our knowledge, Refs. \cite{Chen:2017dsy,Liu:2024tqe} represent the only two existing attempts in this direction. Nevertheless, both of these models lack certain fundamental ingredients.

More precisely, the model in \cite{Liu:2024tqe} incorporates both an axion field, explicitly breaking translational symmetry, and a dilaton field, modifying the infrared structure of the theory. Neither of these elements is present in the MC model and, in our view, should not be required in its holographic dual. Moreover, the holographic model in question can be mapped into the MC framework only at zero charge density. On the other hand, the holographic probe brane model \cite{Chen:2017dsy} correctly captures the MC dynamics but it does not feature an independent scale controlling the MC relaxation time $\tau$, limiting its applicability to the low-temperature regime only.

In this work, we provide a simple holographic dual of the MC model and we verify with numerical and analytical techniques their equivalence. This will also enable us to investigate the fate of chemical shift symmetry and Kubo-Martin-Schwinger (KMS) symmetry in systems featuring non-hydrodynamic modes. Contrary to earlier speculations~\cite{Jain:2023obu,Liu:2018kfw}, our analysis demonstrates that the canonical form of KMS symmetry does not hold for non-hydrodynamic modes. Instead, a new, generalized realization of KMS symmetry emerges.

\section{Maxwell-Cattaneo model}
We consider a relativistic system with U(1) global symmetry and a conserved current $\partial_\mu J^\mu=0$. After defining the four-velocity $u^\mu$, with proper normalization $u_\mu u^\mu=-1$, the current can be parameterized as $J^\mu=n u^\mu+\mathcal{J}^\mu$ where $n$ is the charge density and $\mathcal{J}$ the corresponding flux. The flux obeys the following equation \cite{Jain:2023obu}
\begin{equation}
    \mathcal{J}^\mu+ \tau \Delta^{\mu\nu}u^\lambda \partial_\lambda \mathcal{J}^\nu= - D \Delta^{\mu\nu}\partial_\nu n+\dots \label{ee}
\end{equation}
where external sources have been set to vanish and nonlinear terms have been neglected. Here, $\Delta^{\mu\nu}\equiv \eta^{\mu\nu}+u^\mu u^\nu$, is the spatial projection operator. Moreover, the relation between the charge density $n$ and the chemical potential $\mu$, $\chi\equiv \partial n/\partial \mu$ with $\chi$ the charge susceptibility, has been used. In the limit of $\tau \rightarrow 0$, Eq.~\eqref{ee} reduces to the standard hydrodynamic constitutive equation \cite{Kovtun:2012rj}.

Within this model, the retarded correlation functions of density $n\equiv J^t$ and transverse-flux $J_\perp^i$ can be easily computed \cite{Jain:2023obu},
\begin{align}
    & G^R_{nn}(\omega,k)=\frac{- D \chi k^2}{- {\ri} \omega(1-  {\ri} \omega \tau)-D k^2},\label{G1}\\
&G^R_{J_\perp^iJ_\perp^j}=\frac{ {\ri} \omega D \chi}{1-  {\ri} \omega \tau}\left(\delta^{ij}-\frac{k^i k^j}{k^2}\right),\label{G2}
\end{align}
where $\omega,k$ are respectively the frequency and the wave-vector. From the above equations, the low-energy excitations in the two sectors are given by:
\begin{equation}\label{modes}
    -  {\ri} \omega(1-  {\ri} \omega \tau)-D k^2=0,\qquad \omega=- {\ri}/\tau,
\end{equation}
with the telegrapher equation governing the fluctuation of the charge density $n$.

The MC model just outlined can be formalized in field theory language using the Schwinger-Keldysh (SK) formalism \cite{Liu:2018kfw}. This is achieved by writing down a SK effective action $S_{\text{eff}}$ that is a functional of the gauge-invariant combinations $B_{a\mu}\equiv \partial_\mu \varphi_a+ \mathcal A_{a\mu}$, $B_{r\mu}\equiv \partial_\mu \varphi_r+ \mathcal A_{r\mu}$. Here, we have introduced two sets of Stueckelberg fields $\varphi_a, \varphi_r$ and the corresponding external gauge fields $\mathcal A_{a\mu},\, \mathcal A_{r\mu}$. Despite this is enough in the case of pure diffusion \cite{Liu:2018kfw}, for the MC model, additional spatial vector fields $V_{ai}$, $V_{ri}$ have to be integrated in. Skipping all details, an choosing $u^\mu=(1,0,0,0)$, the SK effective Lagrangian for the MC model is given by (see \cite{Jain:2023obu,Liu:2024tqe} for details)
\begin{align}
    & \mathcal{L}_{\text{eff}}[\varphi_a, \varphi_r, V_{ai}, V_{ri}] \nonumber \\
    &= \chi B_{a0}B_{r0}- \chi_V V_{ai}V_{ri} + {\ri}  T \sigma_0 C_{ai}^2-\sigma_0 C_{ai} \partial_0 C_{ri}\,,\label{MCeft}
\end{align}
with $C_{ai}\equiv B_{ai}+ \chi_V \gamma\, V_{ai}$ and similarly for $C_{ri}$. In the limit $\chi_V \rightarrow 0$, Eq.~\eqref{MCeft} reduces to the standard action for diffusion \cite{Liu:2018kfw}. In general, this action corresponds to the MC model with $\tau=\sigma_0 \gamma^2/\chi_V$.

\section{Holographic dual}
In the gravitational side, we consider the following five-dimensional action
\begin{align}\label{action}
    S =  \int d^{5} x \sqrt{-g} &\left[\frac{1}{2\kappa^2}\left(R+\frac{6}{L^2}\right) -\frac{1}{4e^2} F_{MN}F^{MN} \right.\notag \\ &\qquad \qquad \left.+ \frac{\alpha}{e^2} \left(\frac{1}{4} F_{MN}F^{MN}\right)^2 \right],
\end{align}
where $F=dA$ is the field strength of a bulk gauge field $A_M$, and we assume $\alpha>0$. In the limit $\alpha \rightarrow 0$, \eqref{action} reduces to the standard Maxwell bulk action. Moreover, $2\kappa^2=16\pi G_N$, and $e$ denote the gravitational constant and the Maxwell coupling, respectively. We work in the probe limit, $2\kappa^2/e^2\ll1$, so that the backreaction of the gauge field on the metric $g_{\mu\nu}$ remains negligible. The background metric is then fixed to
\begin{equation}
	d s^{2} = \frac{L^{2}}{u^{2}}\left( -f(u) d t^{2} + d\vec{x}^{2} +\frac{d u^{2}}{f(u)} \right),
\end{equation}
where $(t,\vec{x})$ are (3+1)-dimensional spacetime coordinates in the dual field theory and $u$ is the radial coordinate. Here, $f(u)=1-u^4/u_h^4$ and we set $L=1$. The Hawking temperature is $T=1/(\pi u_h)$.

The equations of motion for the bulk gauge field are given by
\begin{equation}
\label{eom}
    \nabla_M\left[\left(1-\frac{\alpha}{2}F_{PQ}F^{PQ}\right)F^{MN}\right]=0\,,
\end{equation}
and the one point function $\langle J^{\mu} \rangle$ reads
\begin{equation}
\label{Jmu}
\langle J^{\mu} \rangle = \lim_{ u\rightarrow 0}\frac{\delta S_{\rm Gra}}{\delta \mathcal A_\mu} = \lim_{u \rightarrow 0 }\sqrt{-g}\left(1-\frac{\alpha}{2}F_{PQ}F^{PQ}\right)F^{r \mu}\,. 
\end{equation}
Here, $\langle J^t \rangle, \langle J^i \rangle$ are  identified respectively with the charge density and the spatial current in the boundary dual field theory. For more details, we refer to the excellent textbooks \cite{Ammon_Erdmenger_2015, Zaanen_Liu_Sun_Schalm_2015,Natsuume:2014sfa, Baggioli:2019rrs, Hartnoll:2016apf} and lectures \cite{Hartnoll:2009sz} existing in the literature.

The background for the bulk gauge field $A=A_t(u)dt$ satisfies a nonlinear equation
\begin{equation}
\begin{split}
&\left[ u^{-1} A_{t}'(u) + u^{3}\alpha A_{t}'(u)^{3} \right]^\prime = 0 \\ &\qquad \Longrightarrow  A_{t}'(u) + u^{4}\alpha A_{t}'(u)^{3} =-\rho u,\label{vv}
\end{split}
\end{equation}
where $'\equiv d/du$. In obtaining the second line in \eqref{vv}, an integration over $u$ is performed and the integration constant is identified as the background charge density $\langle J^t \rangle \equiv \rho$. This identification becomes obvious by checking the asymptotic solution for $A_t$ near the UV boundary ($u=0$)
\begin{equation}
    A_t(u \to 0)  = \mu - \frac{1}{2}\rho u^2+\dots
\end{equation}
where $\mu$ is the chemical potential in the dual field theory. Here, the holographic dictionary \eqref{Jmu} has been utilized.

From Eq.~\eqref{vv}, there are three solutions to $A_t'(u)$. We are interested in the one regular at the UV boundary $u=0$, which is implicitly given by
\begin{equation}
\label{eq:backsol}
    A_{t}'(u)=\frac{\sqrt[3]{2} \left(A^{\text{aux}}(u)\right)^{2/3}-2 \sqrt[3]{3} \alpha  u^2}{6^{2/3} \alpha  u^3 \sqrt[3]{A^{\text{aux}}(u)}}.
\end{equation}
where we define an auxiliary function as
\begin{equation}
    A^{\text{aux}}(u)\equiv\sqrt{3} \sqrt{\alpha ^3 u^6 \left(27 \alpha  \rho ^2 u^6+4\right)}+9 \alpha ^2 \rho  u^6\,.
\end{equation}

On top of this background solution, we consider the following fluctuations:
\begin{equation}
    A_{\mu}(u,t,x)= A_{\mu}(u) + \delta A_{\mu}(u) e^{-{\ri} \omega t + {\ri} k x}.
\end{equation}
For convenience, we introduce the gauge-invariant longitudinal field as
\begin{equation}
    E_{L}(u)\equiv k \delta A_{t}(u) + \omega \delta A_{x}(u),
\end{equation}
and obtain the equation of motion
\begin{widetext}
\begin{equation}
    E_{L}'' + \frac{E_{L}'}{\omega^{2}- f G k^{2}}\left[ \frac{f'}{f} - \frac{1}{u}\bigg\{ \omega^{2}\big(1+u^4\alpha A_{t}'(3A_{t}'+2uA_{t}'') \big)  - f G k^{2} \frac{1+3u^4\alpha A_{t}'(3A_{t}'+2uA_{t}'')}{1-3u^{4}\alpha A_{t}'^{2}} \bigg\} \right] +\frac{\omega^{2}- f G k^{2}}{f^{2}}E_{L}=0,
\end{equation}
\end{widetext}
where
\begin{equation}
    G = \frac{1-u^{4}\alpha A_{t}'^{2}}{1-3u^{4}\alpha A_{t}'^{2}}.
\end{equation}
Since we are interested in the retarded Green's functions for the dual current operator, we impose the ingoing boundary condition at the black hole horizon as
\begin{equation}
    E_{L}(u) = (u-u_h)^{-{\ri}\frac{\omega}{4\pi T}}{E}_{\rm reg}(u),
\end{equation}
where $E_{\rm reg}(u)$ is a regular function at $u=u_h$. Solving the equation of motion with the ingoing boundary condition, we read off the leading and sub-leading coefficients in the expansion near the AdS boundary ($u=0$),
\begin{equation}
    E_{L}(u) = E_{L}^{(0)}(\omega,k) + E_{L}^{(2)}(\omega,k) u^{2} + \cdots,
\end{equation}
and compute the retarded Green's function for the charge density  \cite{GiuseppePolicastro_2002},
\begin{equation}
    G_{nn}^{R}(\omega,k) = \frac{2k^{2}}{\omega^{2}-k^{2}} \frac{E_{L}^{(2)}(\omega,k)}{E_{L}^{(0)}(\omega,k)}.
\end{equation}
Note that the retarded Green's function for the current is given by $G_{JJ}^{R} = (\omega^{2}/k^{2})G_{nn}^{R}$.
The QNMs can be extracted by computing the frequencies at which the leading coefficient vanishes, $E_{L}^{(0)}=0$, corresponding to the poles of the retarded Green's function. Finally, the optical conductivity is given by
\begin{equation}
    \sigma(\omega) = \frac{G_{JJ}^{R}(\omega,k=0)}{ {\ri} \omega}\,,
\end{equation}
and the DC conductivity is defined as
\begin{equation}
\sigma_0 \equiv \lim_{\omega \to0} {\rm Re}[\sigma(\omega)]\,.
\end{equation}

\begin{figure}[ht]
    \centering
    \includegraphics[width=0.95\linewidth]{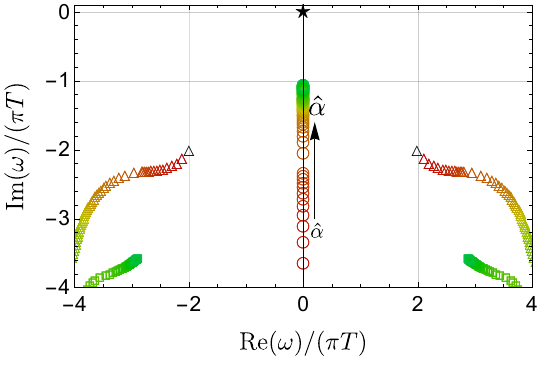}
    \caption{Quasinormal modes in the longitudinal sector at zero wave-vector $k=0$ upon dialing the dimensionless parameter $\hat{\alpha}\equiv \alpha \rho^2$ from $\hat{\alpha}=0$ (darker color) to $\hat{\alpha}\approx31$ (green color). The black star $\bigstar$ indicates the hydrodynamic diffusive mode.}
    \label{fig1}
\end{figure}

\subsection*{Derivation of the analytical formula for the electric DC conductivity}
\label{sec: relaxationtime}
We follow Ref.~\cite{Donos:2015gia} and compute the DC conductivity in our holographic setup. The $x$-component of the current is
\begin{equation}
    \langle J^x\rangle=\lim_{u\rightarrow0}\sqrt{-g}\left(1-\alpha F^2\right)F^{rx}\,.
\end{equation}
Consider the following fluctuation on top of our background fields,
\begin{equation}
\delta A_x = \delta a_{x}(u)-E t\,,
\end{equation}
where $E$ is the $x$-component of the electric field. At the linear level, this generates an electric current
\begin{equation}
\begin{split}
\label{Jx}
\delta \langle J^x \rangle=\frac{f(u)}{u}\delta a_x'(u) \left(1+\alpha u^4 A_t'(u)^2\right)\,,
\end{split}
\end{equation}
that can be computed anywhere in the bulk since~\eqref{Jx} is radially conserved. For convenience, we then evaluate~\eqref{Jx} at the black brane horizon. Near the horizon, the fluctuation $\delta a_x$ is regular in incoming Eddington-Finkelstein coordinates if it behaves as
\begin{equation}
\delta a_x(u)= -\frac{E}{4\pi T}\ln(u-u_h)+\cdots \,.
\end{equation}
After substituting the near horizon behavior into Eq.~\eqref{Jx}, the electric current is expressed as
\begin{equation}
\delta \langle J^x \rangle= E\left(u_h^{-1}+\alpha u_h^4 A_t'(u_h)^2\right)\,.
\end{equation}
Then, using Ohm's law, the electric DC conductivity is simply
\begin{equation}
\label{appB:sigmaDC}
\sigma_0=u_h^{-1}+\alpha u_h^4 A_t'(u_h)^2\,.
\end{equation}
The DC conductivity displays the following asymptotic behaviors
\begin{equation}
\lim_{\hat{\alpha}\rightarrow 0}\frac{\sigma_0}{\pi T}=1\,,
\qquad \lim_{\hat{\alpha}\rightarrow \infty}\frac{\sigma_0}{\pi T}\propto \hat{\alpha}^{1/3},
\end{equation}
where $\hat{\alpha}\equiv \alpha \rho^2$.

\section{Towards the quasi-hydrodynamic regime} 
We now proceed with a numerical study of the dynamics emerging from our holographic model. In Fig.~\ref{fig1}, we show the lowest QNMs in the longitudinal spectrum at zero wave-vector, $k=0$, upon dialing the dimensionless parameter $\hat \alpha \equiv \alpha \rho^2$. At any value of $\hat \alpha$, we observe a hydrodynamic mode $\omega=0$ ($\bigstar$ symbol) at the origin of the complex plane. This is the diffusive mode emerging from the conservation of the U(1) current. In the $\hat \alpha=0$ limit (darker color), that coincides with the Maxwell bulk action usually considered in holography, the lowest non-hydro excitations are a pair of complex modes with frequency $\omega=-({\ri} \pm 1) 2\pi T$. These modes are not well-separated from the other non-hydro modes, forming a so-called \textit{Christmas tree}. In this limit, the description of the dual field theory is not provided by the MC model and does not fit in any quasi-hydrodynamic framework.

The situation changes when $\hat \alpha$ is increased (towards green color in Fig.~\ref{fig1}). More precisely, the pair of complex modes moves down into the complex plane while a purely imaginary mode $\omega=-{\ri}/\tau$ climbs along the imaginary axes. At a critical value of $\hat \alpha$, this mode becomes the lowest non-hydro mode, it gets close to the hydro one and parametrically separated from the other non-hydro modes. This defines the regime where quasi-hydrodynamics applies and where the dual dynamics are described by the MC model.

\subsection{More details on the regime of validity of the MC model}
The MC model relies on two simplifying assumptions: (i) only one non-hydrodynamic mode is relevant, and (ii) higher-order gradient corrections can be neglected. These assumptions break down when (i) the frequency is high enough for additional non-hydro modes to become relevant, and (ii) the wave-vector is large enough so that higher-order gradients can no longer be ignored.

\begin{figure}[h]
    \centering
    \includegraphics[width=0.8\linewidth]{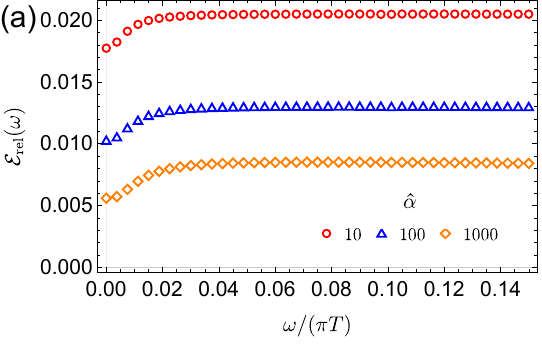}

    \vspace{0.2cm}
    
     \includegraphics[width=0.8\linewidth]{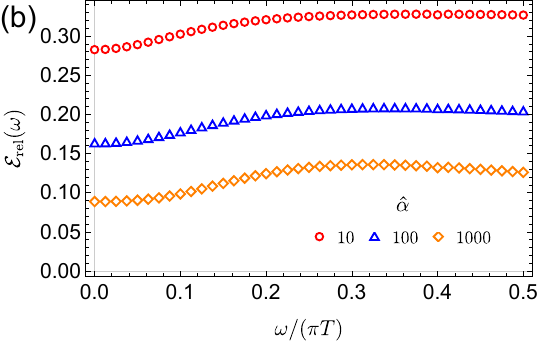}
    \caption{Relative error $\mathcal{E}_{\text{rel}}(\omega)\equiv |G^R_{nn,\text{numeric}}(\omega)-G^R_{nn,\text{MC}}(\omega)|/G^R_{nn,\text{MC}}(0)$ as a function of $\omega$ for different values of $\hat{\alpha}$. \textbf{(a)} $k/(\pi T)=0.1$, \textbf{(b)} $k/(\pi T)=0.4$.}
    \label{fig:end1}
\end{figure}

To verify these limitations, we compare the predictions of the MC model for the density-density correlator, Eq.~\eqref{G1}, with numerical data. In Fig.~\ref{fig:end1}(a), we show this comparison for small $k$, where gradient corrections are suppressed. We find that increasing $\hat{\alpha}$ improves the agreement between model and data. This trend is consistent with the spectra shown in Fig.~\ref{fig1}, where larger $\hat{\alpha}$ shifts the first non-hydro mode (included in the MC model) closer to the origin, while pushing the remaining non-hydro modes deeper into the complex frequency plane.

The other type of corrections, originating from spatial gradients, become apparent when performing the same comparison at larger values of $k$, as shown in Fig.~\ref{fig:end1}(b). In this regime, the discrepancies are significantly larger, though they still decrease in the large $\hat{\alpha}$ limit. Importantly, improving the agreement here does not require including additional non-hydrodynamic modes, but rather incorporating higher-order spatial gradient terms into the quasi-hydrodynamic framework.

\begin{figure}[ht]
    \centering\includegraphics[width=0.85\linewidth]{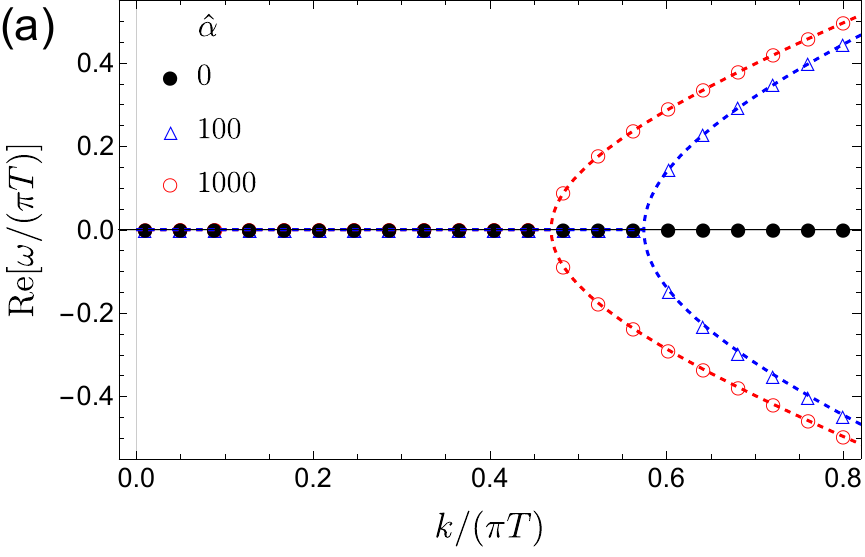}

      \vspace{0.2cm}

    \includegraphics[width=0.85\linewidth]{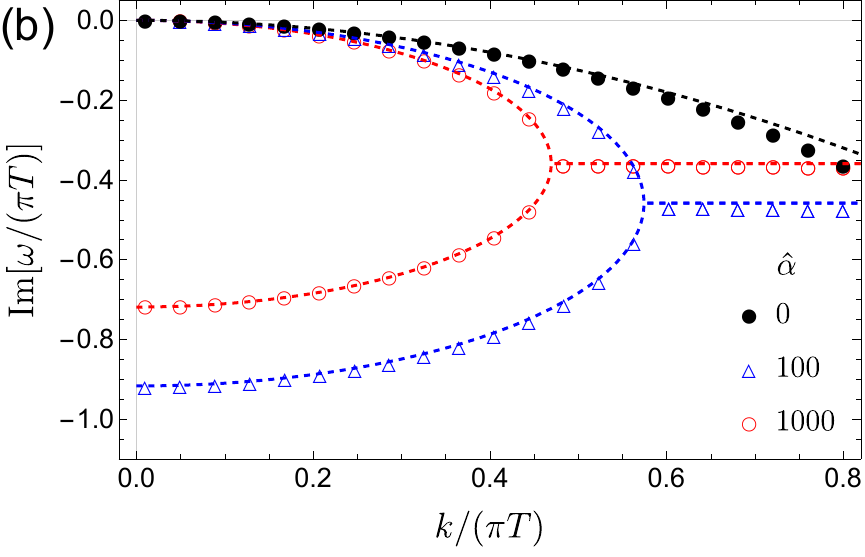}
    \caption{Dispersion relation $\omega(k)$ of the lowest QNMs for three different values of $\hat \alpha$. Panels \textbf{(a)}-\textbf{(b)} refer respectively to the real and imaginary part of the frequency. Symbols are the results of the numerical computation while dashed lines are the fits to the theoretical dispersion obtained from Eq.~\eqref{mo}.}
    \label{fig2}
\end{figure}

\section{Numerical \& analytical tests of the equivalence} 
In Fig.~\ref{fig2}, we present the dispersion of the lowest QNMs as a function of the dimensionless wave-vector $k/(\pi T)$ for three different values of $\hat \alpha$. At $\hat{\alpha}=0$ (black symbols), we observe the expected diffusive behavior $\omega=-{\ri} D k^2 +\dots$ where $D=\sigma_0/\chi$. On the other hand, at finite (and large enough) $\hat \alpha$, the dynamics involve one hydrodynamic diffusive mode and a second non-hydro mode that collide in the complex plane at a critical wave-vector, inducing a propagating wave-like excitation at large wave-vector. These dynamics are well captured by the telegrapher equation,
\begin{equation}
    \omega^2+{\ri} \omega /\tau = v_0^2 k^2,\qquad v_0^2=D/\tau,\label{mo}
\end{equation}
whose solutions are displayed with dashed lines in Fig.~\ref{fig2}, confirming the validity of the MC model also at $k\neq 0$.

\subsection{Derivation of the analytical formula for the relaxation time} \label{tau-match_method}
Aside from directly fitting the numerical results for the QNMs with the telegrapher equation~\eqref{mo}, a more sophisticated derivation of the relaxation time may be possible using a matching method.

Consider the following fluctuation ansatz
\begin{equation}
\label{ansatz:fluctuation}
A=\left(A_t(u)+\delta A_t(u,x^\mu)\right)dt+\delta A_x(u,x^u) dx\,.
\end{equation}
The equations of motion for the fluctuations can be written as
\begin{subequations}
\label{Eq:fluctuation}
\begin{equation}
\partial_u\left(X_3\delta A_t'\right)+\frac{k X_1}{f}\left(\omega \delta A_x + k \delta A_t\right)=0\,,\label{Eq:fluctuationa}
\end{equation}
\begin{equation}
\partial_u\left(f X_5 \delta A_x'\right)-\frac{\omega X_1}{f}\left(\omega \delta A_x + k \delta A_t\right)=0\,,\label{Eq:fluctuationb}
\end{equation}
\begin{equation}
\omega X_3 \delta A_t' + k f X_5 \delta A_x' = 0\,,\label{Eq:fluctuationc}
\end{equation}
\end{subequations}
where we take $\partial_t \rightarrow -{\ri}\omega$, $\partial_x\rightarrow {\ri} k$, and define
\begin{equation}
\begin{split}
&X_1=-\frac{L}{u} \left(1+\alpha\frac{ u^{4}}{L^4}A_t'(u)^2\right)\,,\\ 
&X_3=-\frac{L}{u} \left(1+3\alpha\frac{ u^{4}}{L^4}A_t'(u)^2\right)\,,\\
&X_5=\frac{L}{u} \left(1+\alpha\frac{ u^{4}}{L^4}A_t'(u)^2\right)\,.
\end{split}
\end{equation}
When $\omega\rightarrow 0, k \rightarrow 0$, in the region near the AdS boundary, the solution to Eq.~\eqref{Eq:fluctuation} takes the following form
\begin{equation}
\label{sol:outer}
\begin{split}
    &\delta A_t(u,x^\mu) = \delta a_t(x^\mu)+\delta j_t(x^\mu)\Phi_1(u)+\mathcal{O}(\omega,k)\,,\\
    &\delta A_x(u,x^\mu) = \delta a_x(x^\mu)+\delta j_x(x^\mu)\Phi_2(u)+\mathcal{O}(\omega,k)\,,
\end{split}
\end{equation}
where
\begin{equation}
\Phi_1=\int_0^u ds\frac{1}{X_3(s)}\,,\quad \Phi_2=\int_0^u ds\frac{1}{f(s)X_5(s)}\,.
\end{equation}
After substituting Eq.~\eqref{sol:outer} into Eq.~\eqref{Eq:fluctuationc}, we obtain the conservation equation
\begin{equation}
\label{Eq:consevation}
\omega \delta j_t + k \delta j_x =0\,.
\end{equation}
Note that, near the boundary, $\delta A_x$ in Eq.~\eqref{sol:outer} has a logarithmic divergence because the integrand of $\Phi_2$ diverges as $u\rightarrow u_h$, while $\delta A_t$ is regular. We rewrite Eq.~\eqref{sol:outer} as
\begin{equation}
\label{sol:outer2}
\begin{split}
    \delta A_t &\approx \delta a_t(x^\mu)+\delta j_t(x^\mu)\phi_1(u)\,,\\
    \delta A_x &\approx \delta a_x(x^\mu)+\delta j_x(x^\mu)\phi_2(u) \\
    &+\frac{\delta j_x(x^\mu)\ln(f(u))}{f'(u_h)X_5(u_h)}\,,
\end{split}
\end{equation}
where we introduced regular functions
\begin{equation}
\begin{split}
    \phi_1(u)&\equiv \Phi_1(u)\,,\\
    \phi_2(u)&\equiv \Phi_1(u)-\frac{\ln(f(u))}{f'(u_h)X_5(u_h)} \\
    &=\int^{u}_0 ds\frac{1}{f(s)X_5(s)}\left(1-\frac{f'(s)X_5(s)}{f'(u_h)X_5(u_h)}\right)\,.
\end{split}
\end{equation}
On the other hand, near the black hole horizon, the solution is
\begin{equation}
\label{sol:inner}
\begin{split}
&\delta A_t=\delta a_{t}^{(H,1)}(u,x^\mu)+\delta a_{t}^{(H,2)}(u,x^\mu)e^{-\frac{{\ri}\omega}{4\pi T}\ln(f(u))}\,,\\
&\delta A_x=\delta a_{x}^{(H,1)}(u,x^\mu)+\delta a_{x}^{(H,2)}(u,x^\mu)e^{-\frac{{\ri}\omega}{4\pi T}\ln(f(u))}\,,
\end{split}
\end{equation}
where $\delta a_{\mu}^{(H,1)}$ is a pure gauge solution, and $\delta a_{\mu}^{(H,2)}$ is an incoming solution with $\mu=t,x$. Near the black hole horizon, the incoming solution has a logarithmic behavior
\begin{equation}
\begin{split}
e^{-\frac{{\ri}\omega}{4\pi T}\ln(f(u))}&\approx 1 -\frac{{\ri}\omega}{4\pi T}\ln(f(u))+\mathcal{O}(\omega^2) \\
&=1+\frac{\ln(f(u))}{4\pi T}\partial_t+\ldots\,. 
\end{split}
\end{equation}
We rewrite Eq.~\eqref{sol:inner} as
\begin{equation}
\label{sol:inner2}
\begin{split}
&\delta A_t\approx\delta a_{t}^{(H,1)}+\delta a_{t}^{(H,2)}+\frac{\ln(f(u))}{4\pi T}\partial_t \delta a_{t}^{(H,2)}\,,\\
&\delta A_x\approx\delta a_{x}^{(H,1)}+\delta a_{x}^{(H,2)}+\frac{\ln(f(u))}{4\pi T}\partial_t a_{x}^{(H,2)}\,.
\end{split}
\end{equation}
In the intermediate region, we match regular the terms and logarithmic terms from Eq.~\eqref{sol:outer2} and Eq.~\eqref{sol:inner2} separately:
\begin{equation}
\begin{split}
&\delta a_{t}^{(H,1)}(u,x^\mu)+\delta a_{t}^{(H,2)}(u,x^\mu)=\delta a_t(x^\mu)  \\
&+\delta j_t(x^\mu)\phi_1(u)\,,\\
&\delta a_{t}^{(H,2)}(u,x^\mu)=0\,,\\
&\delta a_{x}^{(H,1)}(u,x^\mu)+\delta a_{x}^{(H,2)}(u,x^\mu)=\delta a_x(x^\mu)  \\
& +\delta j_x(x^\mu)\phi_2(u)\,,\\
&\partial_t\delta a_{x}^{(H,2)}=\frac{4\pi T}{f'(u_h)X_5(u_h)}\delta j_x(x^\mu)\,.\\
\end{split}
\end{equation}
By taking the appropriate derivative and linear combination, we arrive at
\begin{equation}
\begin{split}
\partial_t \delta j_x - \frac{\phi_1(u_h)}{\phi_2(u_h)} \partial_x \delta j_t&=-\frac{1}{\phi_2(u_h)}\left(\partial_t \delta a_x-\partial_x \delta a_t\right)\\
&-\frac{1}{\phi_2(u_h)X_5(u_h)}\delta j_x ,   
\end{split}
\end{equation}
with the pure gauge solution satisfying
\begin{equation}
    \partial_x \delta a_{t}^{(H,1)}-\partial_t \delta a_{x}^{(H,1)}=0\,.
\end{equation}
Using the conservation equation, Eq.~\eqref{Eq:consevation}, and the following identifications
\begin{equation}
\label{eq:tau}
v_0^2=\frac{\phi_1(u_h)}{\phi_2(u_h)}\,,\quad \chi=\frac{1}{\phi_2(u_h)}\,,\quad
\tau=X_5(u_h)\phi_2(u_h)\,,
\end{equation}
we obtain the final equations for the fluctuations:
\begin{equation}
\begin{split}
&\partial_t \delta j^t + \partial_x\delta j^x =0\,,\\
&\partial_t \delta j^x + v^2 \partial_x \delta j^t=-\chi\left(\partial_t \delta a_x-\partial_x \delta a_t\right)-\frac{1}{
\tau
}\delta j^x\,,
\end{split}\label{bubu}
\end{equation}
where $\partial_t \delta a_x=E_x$ and $\partial_x \delta a_t= \partial_x \delta \mu$.

We notice that Eqs.~\eqref{bubu} match exactly the dynamics expected from the Maxwell-Cattaneo model with relaxation time $\tau$ and diffusion constant $D= v^2 \tau$. This is the first hint that our holographic model captures the correct dynamics and is, in this sense, a simple gravitational dual of the MC model.
\begin{figure}
    \centering
    \includegraphics[width=0.85\linewidth]{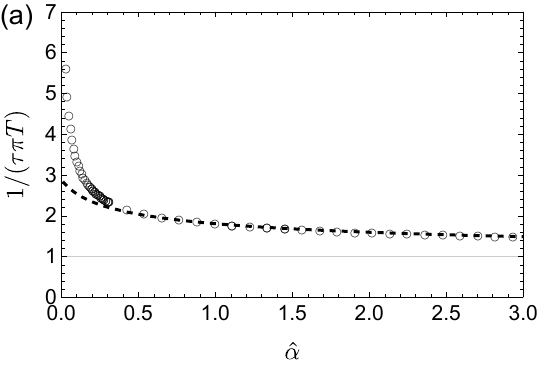}

    \vspace{0.2cm}

    \includegraphics[width=0.85\linewidth]{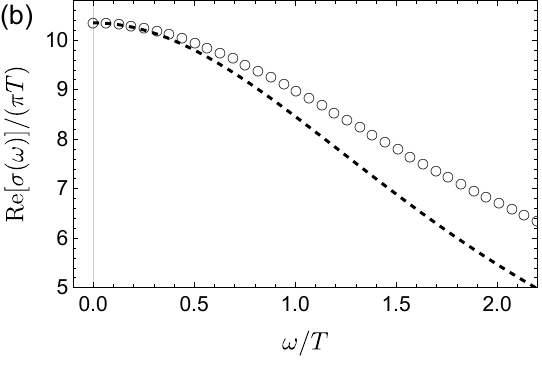}

    \caption{\textbf{(a)} The relaxation time extracted from the imaginary part of the imaginary QNM (open circles in Fig.~\ref{fig1}) compared with the theoretical prediction (dashed line) from Eq.~\eqref{tautheory} as a function of $\hat \alpha$. \textbf{(b)} The real part of the optical conductivity $\mathrm{Re}[\sigma(\omega)]$ as a function of frequency with $\hat{\alpha}=1000$. The symbols are the numerical data and the dashed lines the theoretical predictions using the Drude formula with the relaxation time defined in Eq.~\eqref{tautheory} and the DC conductivity taken from Eq.~\eqref{appB:sigmaDC}.}
    \label{fig3}
\end{figure}
\subsection{Testing the analytical predictions}
The previous analysis yields an analytic prediction for the relaxation time $\tau$, that is valid within the quasi-hydrodynamic limit,
\begin{equation}
    \tau=X(u_h)Y(u_h), \label{tautheory}
\end{equation}
with:
\begin{align}
  &Y(u)=\int^{u}_0 ds\frac{1}{f(s)X(s)}\left(1-\frac{f'(s)X(s)}{f'(u_h)X(u_h)}\right)',\nonumber\\
  &X(u)\equiv \frac{1}{u}+\alpha u^3 A_t'(u)^2.
\end{align}
In Fig.~\ref{fig3}(a), we compare the relaxation time extracted from the imaginary part of the purely imaginary QNM $\omega=-{\ri}/\tau$ (open circles in Fig.~\ref{fig1}) with our theoretical prediction, Eq.~\eqref{tautheory}. The two are in good agreement with each other when $\tau^{-1}\gtrapprox 2\pi T$. This coincides exactly with the value of $\hat \alpha$ at which this mode becomes the lowest non-hydro mode and the quasi-hydrodynamic approximation becomes valid.

To further confirm this fact, in Fig.~\ref{fig3}(b) we show the real part of the optical conductivity $\sigma(\omega)$ with $\hat \alpha = 1000$. In the low frequency region, the conductivity is well-fitted by a Drude form $\sigma(\omega)= \sigma_0/(1-{\ri} \omega \tau)$ where $\sigma_0$ is obtained from Eq.~\eqref{appB:sigmaDC}.

Moreover, the Einstein relation $\sigma_0=D \chi$ holds. We note that the agreement between the numerical data and the Drude model improves progressively with increasing $\hat{\alpha}$, as the higher-order non-hydrodynamic modes (see Fig.~\ref{fig1}), which are neglected in the Maxwell-Cattaneo model, shift deeper into the complex plane.

\begin{figure}
    \centering
    \includegraphics[width=0.85\linewidth]{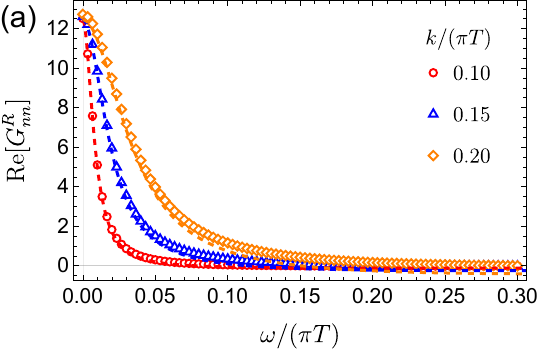}

    \vspace{0.2cm}

    \includegraphics[width=0.85\linewidth]{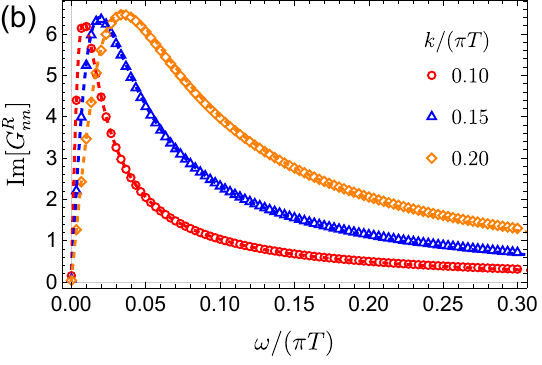}
    \caption{Charge density retarded correlator $G^R_{nn}(\omega,k)$ as a function of $\omega/(\pi T)$ with $\hat\alpha=1000$ and several values of $k/(\pi T)$. The dashed lines denote the analytic form, Eqs.~\eqref{G1}-\eqref{G2}, with $\sigma_0=D \chi$ and $\tau$ analytically computed.}
    \label{fig4}
\end{figure}

Finally, in Fig.~\ref{fig4}, we show the retarded correlator for the charge density operator as a function of frequency and for finite wave-vector. The numerical data match the theoretical formulae, Eqs.~\eqref{G1}-\eqref{G2} from the Maxwell-Cattaneo model. This confirms that our holographic model is dual to the MC model.
\section{Holographic Schwinger-Keldysh effective action}
To further validate the equivalence between our minimal holographic model and the Maxwell-Cattaneo (MC) framework, we employ holographic Schwinger-Keldysh (holoSK) techniques \cite{Glorioso:2018mmw, Crossley:2015tka, haehl2016fluid}, which have recently been used to study quasi-hydrodynamic systems \cite{Baggioli:2023tlc} and extended to incorporate non-hydrodynamic modes \cite{Liu:2024tqe}.

We introduce a new methodology for incorporating non-hydrodynamic modes into the holoSK formalism, based on constructing bulk solutions localized at specific characteristic frequencies (see \cite{Sakai:2004cn} for a similar procedure in the context of holographic QCD). A detailed description of this novel approach, together with the corresponding technical derivations, is provided in the following sections.

Using this framework, we analytically derive the boundary Schwinger-Keldysh action corresponding to our holographic model, Eq.~\eqref{action}, and demonstrate that it precisely matches the effective field theory description of the Maxwell-Cattaneo model proposed in \cite{Jain:2023obu}. 

We note that an alternative method for incorporating non-hydrodynamic modes into the holoSK framework was recently proposed in \cite{Liu:2024tqe}. We comment in more detail on the differences between the two methods in Section~\ref{compi}.

\subsection{Technical derivation}
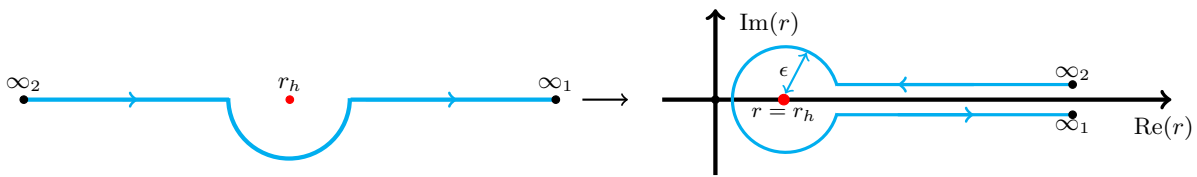
\begin{figure} [htbp!]
		\centering
	\begin{tikzpicture}[]
		\draw[cyan, ultra thick] (-3.5,0)--(-0.8,0);
		\draw[cyan, ultra thick] (0.8,0)--(3.5,0);
		\draw[cyan, ultra thick] (-0.81,0.019) arc (-180:0:0.8);
		\draw[cyan, ->,very thick] (1,0)--(2.2,0);
		\draw[cyan, ->,very thick] (-2.2,0)--(-2,0);
		\draw[fill] (-3.5,0) circle [radius=0.05];
		\node[above] at (-3.5,0) {\small $\infty_2$};
		\draw[fill] (3.5,0) circle [radius=0.05];
		\node[above] at (3.5,0) {\small $\infty_1$};
		\draw[fill,red] (0,0) circle [radius=0.05];
		\node[above] at (0,0) {\small $r_h$};				
		
		\draw[-to,thick] (0,-1.15)--(0,-1.75);

		\node[below] at (3.3,-3.1) {\small Re$(r)$};
		\draw[->,ultra thick] (-3.3,-3)--(3.4,-3);
		\node[right] at (-2.4,-2) {\small Im$(r)$};
		\draw[->,ultra thick] (-2.6,-4)--(-2.6,-1.8);
		\draw[cyan, very thick] (-1,-3.18) arc (165:-165:-0.7);
		\draw[cyan, very thick] (-1,-2.8)--(2.1,-2.8);
		\draw[cyan, very thick] (-1,-3.2)--(2.1,-3.2);
		\draw[cyan, ->,very thick] (-0.2,-3.2)--(0.8,-3.2);
		\draw[cyan, <-,very thick] (-0.2,-2.8)--(0.8,-2.8);
		\draw[fill] (-2.6,-3) circle [radius=0.05];
		\draw[fill] (2.1,-2.8) circle [radius=0.05];
		\node[below] at (2.1,-3.15) {\small $\infty_1 $};
		\draw[fill] (2.1,-3.2) circle [radius=0.05];
		\node[above] at (2.1,-2.85) {\small $\infty_2 $};
		\draw[fill , red ] (-1.7,-3) circle [radius=0.07];
		\node[below] at (-1.7,-3) {\footnotesize $r=r_h $};
		\draw[cyan, thick,<->] (-1.68,-2.92)--(-1.4,-2.38);
		\node[above] at (-1.7, -2.8) {\footnotesize $\epsilon$};
	\end{tikzpicture}
		\caption{From complexified double AdS (left) \cite{Crossley:2015tka} to the holographic SK contour (right) \cite{Glorioso:2018mmw}.} \label{holo_contour}
\end{figure}

In order to derive the boundary effective action, we solve the bulk dynamics in the double Schwarzschild-AdS$_5$ spacetime, whose metric in the ingoing Eddington-Finkelstein (iEF) coordinate system is given by
\begin{align}
ds^2 = g_{MN} dx^M dx^N = 2 dr dv - r^2 f(r) dv^2 + r^2 \delta_{ij} dx^i dx^j ,\label{metric_EF} 
\end{align}
where $f(r) = 1- r_h^4/r^4$ and the holographic coordinate $r$ varies along the contour shown in Fig.~\ref{holo_contour}. 
\\
The radial contour in Fig.~\ref{holo_contour} is the gravity dual of the boundary SK closed time path, and has been proven efficient in obtaining the SK effective action. We will always assume the following gauge choice
\begin{align}
A_r = - \frac{A_0}{r^2f(r)} \Longleftrightarrow \tilde A_r = 0\,,
\end{align}
where a tilde is to denote the quantity in the Poincare patch. We split the bulk gauge field into a homogeneous background and a small perturbation
\begin{align}
A_M(r,x^\mu) = \delta^M_0 \bar A_M(r) + \delta A_M(r,x^\mu)\,.
\end{align}
The perturbation $\delta A_M$ satisfies a linear system of partial differential equations,
\begin{align}
    \nabla_M \left[ (1 + \mathcal C)\delta F^{MN} - 2\mathcal C \delta^{N0} \delta^{Mr} \nabla_r \delta A_0 \right]=0\,,\label{PDE_deltaA}
\end{align}
where $\delta F^{MN} = \nabla^M \delta A^N - \nabla^N \delta A^M$ and $\mathcal C = \alpha (\partial_r\bar A_0)^2$. Explicitly, we have
\begin{align}
\mathcal C =& -\frac{2}{3} + \frac{\left(\sqrt{{\hat \rho}^2 + r^6}+ \hat \rho\right)^{2/3}}{3 r^2}+\frac{r^2}{3\left( \sqrt{{\hat \rho}^2 + r^6 }+\hat \rho \right)^{2/3}} \nonumber \\
&{\rm with} \qquad  \hat \rho = \frac{ \sqrt{27 \alpha  \rho^2}}{2}\,.
\end{align}

While each hydro mode satisfies $\omega(k)\to 0$ as $k\to 0$, a non-hydro mode is characterized by $\omega(k) \neq 0 $ as $k\to 0$. With this in mind, we formally write \cite{Sakai:2004cn}

\begin{align}
\delta A_\mu(r,x^\alpha) = \delta A_\mu^{\rm h}(r,x^\alpha) + \delta A_\mu^{\rm nh}(r,x^\alpha) + \cdots \,,\label{Cmu_solution}
\end{align}
where $\delta A_\mu^{\rm h}$ and $\delta A_\mu^{\rm nh}$ correspond respectively to the hydro mode and the leading non-hydro mode. We will see that $\delta A_\mu^{\rm nh}$ is characterized by an intrinsic frequency related to the relaxation time $\tau$. The treatment \eqref{Cmu_solution} represents a major difference from \cite{Liu:2024tqe}, where the non-hydrodynamic
mode is introduced through Dirichlet boundary conditions applied at a stretched horizon. We will adopt a partially on-shell prescription \cite{Crossley:2015tka} when solving Eq.~\eqref{PDE_deltaA}: solving the dynamical components of Eq.~\eqref{PDE_deltaA} to determine $\delta A_\mu$, while leaving aside the constraint component of Eq.~\eqref{PDE_deltaA}.

Near the AdS boundaries, we have
\begin{align}
\delta A_\mu^{\rm h}(r\to \infty_s, x^\alpha) &= B_{s\mu}(x^\alpha) + \cdots \nonumber \\
&+ \frac{\mathfrak J_{s\mu}^{\rm h}(x^\alpha)}{r^2} + \cdots\,,  \nonumber \\
\delta A_\mu^{\rm nh}(r\to \infty_s, x^\alpha) &= \mathcal V_{s\mu}(x^\alpha) + \cdots \nonumber \\
&+ \frac{\mathfrak J_{s\mu}^{\rm nh}(x^\alpha)}{r^2} + \cdots\,. \label{Cmu_boundary}
\end{align}
Here, $B_{s\mu} \equiv \mathcal A_{s\mu} + \partial_\mu \varphi_s$, as used in the main text, Eq. (8). It turns out that $\mathfrak J_{s\mu}^{\rm nh}$ is the suitable dynamical variable for the non-hydro mode. Thus, solving the bulk EOMs yields $\mathfrak J^{\rm h}_{s\mu} = \mathfrak J^{\rm h}_{s\mu}[B_{s\mu}]$ and $\mathcal V_{s\mu} = \mathcal V_{s\mu}[\mathfrak J_{s\mu}^{\rm nh}]$. Imposing the bulk EOMs \eqref{PDE_deltaA} and utilizing Eq. \eqref{Cmu_boundary}, we reduce the quadratic part of the bulk action into a four-dimensional boundary term
\begin{align}
S_0^{(2)} = \int d^4 x & \left[  B_{r\mu} \mathfrak J^\mu_{{\rm h},a} + B_{a\mu} (\mathfrak J^\mu_{{\rm h},r}  + \mathfrak J^\mu_{{\rm nh},r}) \right. \nonumber \\
 & + \mathcal V_{r\mu} \mathfrak J_{{\rm h},a}^\mu + B_{r\mu} \mathfrak J_{{\rm nh},a}^\mu + \mathcal V_{r\mu} \mathfrak J_{{\rm nh},a}^\mu \nonumber \\
 &   + \mathcal V_{a\mu} (\mathfrak J^\mu_{{\rm h},r} + \mathfrak J^\mu_{{\rm nh},r}) \left. \right]\,,\label{B+V_J}
\end{align}
which will give the effective action, Eq. (8) in the main text, once the bulk solutions are obtained. 
 
The hydro solution $\delta A_\mu^{\rm h}$ is constructed perturbatively in the hydrodynamic limit $\partial_\mu \to 0$ (gradient expansion):
\begin{align}
\delta A_\mu^{\rm h} = \delta A_\mu^{{\rm h}\,(0)} + \delta A_\mu^{{\rm h}\,(1)} + \cdots\,,
\end{align}
where the superscript $^{(n)}$ counts the number of boundary derivatives $\partial_\mu$. The leading part is given by
\begin{align}
 \delta A_0^{{\rm h}\,(0)}(r) =& B_{s0} \left( \int_{r_h}^{\infty_s} \frac{dy}{y^3 [1+3 \mathcal C(y)]} \right)^{-1} \nonumber \\
&\int_{r_h}^r \frac{dy}{y^3 [1+3 \mathcal C(y)]}\,, \nonumber \\
 \delta A_i^{{\rm h}\,(0)}(r) =& B_{2i} + B_{ai} \left( \int_{\infty_2}^{\infty_1} \frac{dy}{y^3f(y) [1+ \mathcal C(y)]} \right)^{-1}  \nonumber \\
 &\int_{\infty_2}^r \frac{dy}{y^3f(y) [1+ \mathcal C(y)]}\,.
\end{align}
The solution for $\delta A_0^{\rm h\,(1)}$ takes a piecewise form
\begin{align}
& \delta A_0^{{\rm h}\, (1)}(r) = -\zeta_s(r) \partial_0 \delta A_0^{{\rm h}\,(0)}(r)\,, \nonumber \\
&{\rm with} \quad  \zeta_s(r) = \int_{\infty_s}^r \frac{dy}{y^2f(y)}\,, \quad r\in[r_h - \epsilon, \infty_s)\,.\label{C01_hydro}
\end{align}
The solution for $\delta A_i^{\rm h\, (1)}$ reads
\begin{align}
& \delta A_i^{{\rm h}\, (1)}(r) = \zeta(r) \partial_0 \delta A_i^{{\rm h}\,(0)}(r) \nonumber \\
&+ r_h \left[ 1 + \mathcal C(r_h) \right] \partial_0 B_{1i} \int_{\infty_2}^r \frac{dy}{y^3 f(y) [1 + \mathcal C(y)]}\,, \label{Ci1_hydro}
\end{align} 
where 
\begin{align}
\zeta(r) = \int_{\infty_2}^r \frac{dy}{y^2f(y)}\,, \qquad r\in (\infty_2, \infty_1)\,.
\end{align}
Near the AdS boundaries, we read off the hydro expansion of normalizable modes $\mathfrak J_{s\mu}^{\rm h}$
\begin{align}
\mathfrak J_{s0}^{\rm h} & =  -\frac{1}{2} B_{s0} \left( \int_{r_h}^{\infty_s} \frac{dy}{y^3 [1+3 \mathcal C(y)]} \right)^{-1}\,, \nonumber \\
\mathfrak J_{2i}^{\rm h} & = \frac{{\rm i} r_h^2}{\pi} [1+ \mathcal C(r_h)] B_{ai} - \frac{1}{2} r_h[1+ \mathcal C(r_h)] \partial_0 B_{1i}\,, \nonumber \\
\mathfrak J_{1i}^{\rm h} & = \frac{{\rm i} r_h^2}{\pi} [1+ \mathcal C(r_h)] B_{ai} - \frac{1}{2} r_h[1+ \mathcal C(r_h)] \partial_0 B_{2i}\,. \label{Cmu_h_normalizable}
\end{align}

To solve $\delta A_\mu^{\rm nh}$, we turn to the Fourier space, $\partial_\mu \to (-{\rm i} \omega, {\rm i} \vec k)$. Recall that $\delta A_\mu^{\rm nh}$ is characterized by a {\it finite} frequency even when $\vec k\to 0$. We are thus motivated to expand $\delta A_\mu^{\rm nh}$ in the limit $\vec k \to 0$ without using the $\omega \rightarrow0$ limit:
\begin{align}
\delta A_\mu^{{\rm nh}} = \delta A_\mu^{{\rm nh}\,[0]} + \delta A_\mu^{{\rm nh}\,[1]} + \cdots\,,
\end{align}
where the superscript $^{[n]}$ denotes the power of $\vec k$.

The equation for $\delta A_0^{{\rm nh}\, [0]}$ can be recast in a compact form, 
\begin{align}
&0 = \partial_r\left[ r^3 (1 + 3 \mathcal C)\partial_r \widetilde {\delta A}_0^{{\rm nh}\, [0]}  \right]\,, \nonumber \\
&{\rm with} \qquad \widetilde {\delta A}_0^{{\rm nh}\, [0]}(r,\omega) = e^{-{\rm i} \omega \zeta_s(r)} \delta A_0^{{\rm nh}\, [0]}(r,\omega)\,. \label{C0_nh0_eom}
\end{align}
Thus, the solution for $\delta A_0^{{\rm nh}\, [0]}$ is 
\begin{align}
\delta A_0^{{\rm nh}\, [0]}(r,\omega) &= e^{{\rm i} \omega \zeta_s(r)} \biggr[ \#_1  \nonumber \\
&+ \#_2 \int_{r_h}^r \frac{dy}{y^3[1+ 3 \mathcal C(y)]} \biggr]\,, \label{C0nh_0}
\end{align}
which may be absorbed into the hydro solution $\delta A_\mu^{\rm h}$ via the redefinition of the charge density. This is equivalent to setting $\mathfrak J_{s0}^{\rm nh} = 0$, yielding
\begin{align}
\delta A_0^{{\rm nh}\, [0]}(r,\omega) = 0\,. \label{C0nh_0_zero}
\end{align}
The equation for $\delta A_i^{{\rm nh}\,[0]}$ is
\begin{align}
0& = \partial_r \left[ r^3f(r) (1 + \mathcal C)\partial_r \delta A_i^{{\rm nh}\,[0]} \right] - 2{\rm i} \omega r (1 + \mathcal C)\partial_r \delta A_i^{{\rm nh}\,[0]}\nonumber \\
&- {\rm i} \omega \partial_r\left[ r(1 + \mathcal C) \right] \delta A_i^{{\rm nh}\,[0]} \,,\label{EOM_Cinh_k0_EF}
\end{align}
which is solved by
\begin{align}
\delta A_i^{{\rm nh}\,[0]} (r,\omega) &= \left[ \frac{\mathfrak J_{ri}^{\rm nh}(\omega)}{\mathcal O(\omega)} + \frac{1}{2} {\rm coth}\left( \frac{\beta \omega}{2} \right) \frac{\mathfrak J_{ai}^{\rm nh}(\omega)}{\mathcal O(\omega)} \right] \Phi(r,\omega) \nonumber \\
&+ \frac{1}{e^{-\beta \omega} - 1} \frac{\mathfrak J_{ai}^{\rm nh}(\omega)}{\mathcal O(-\omega)} e^{2{\rm i}\omega \zeta(r)} \Phi(r,-\omega)\,.\label{Cinh_k0_sol}
\end{align}
Here $\Phi(r,\omega)$ represents a {\it regular} solution to Eq. \eqref{EOM_Cinh_k0_EF} that behaves as
\begin{align}
\Phi(r \to \infty,\omega) &= \phi(\omega) - \frac{{\rm i} \omega \phi(\omega) }{r} \nonumber \\
& + \frac{1}{2} \omega^2 \phi(\omega) \frac{\log r}{r^2} + \frac{\mathcal O(\omega) }{r^2} + \cdots\,.
\end{align}
Imposing Dirichlet boundary conditions
\begin{align}
\Phi(r=\infty, \omega) = 0 \Rightarrow \phi(\omega) = 0\,,
\end{align}
this yields a discrete set of 
characteristic frequencies (i.e., quasinormal modes), among which we are interested in the first non-trivial one $\omega_1$, corresponding to the leading non-hydro mode. For general $\omega$, we numerically construct $\Phi$. However, the matching method of Ref. \cite{Grozdanov:2018fic} may be implemented to yield an approximate analytical solution for $\delta A_i^{{\rm nh}\, [0]}$ and thus to obtain an approximate analytical formula for the characteristic frequency $\omega_1$. This procedures reproduces the holographic formula Eq. (13) for $\tau$ reported in the main text. 

We proceed to the next order, $\delta A_\mu^{{\rm nh}\, [1]}$. Leveraging on Eq. \eqref{C0nh_0_zero}, $\delta A_i^{{\rm nh}\, [1]}$ and $\delta A_i^{{\rm nh}\, [0]}$ obey the same equation, which implies that one can set $\delta A_i^{{\rm nh}\, [1]} = 0$. We are left with $\delta A_0^{{\rm nh}\, [1]}$
\begin{align}
0 &= \partial_r\left[ r^3 (1 + 3 \mathcal C)\partial_r \widetilde {\delta A}_0^{{\rm nh}\, [1]}  \right] \nonumber \\
&- \frac{\omega k_i}{r f(r)}(1 + \mathcal C)  \widetilde {\delta A}_i^{{\rm nh}\,[0]}\,,
\end{align}
where $\delta A_0^{{\rm nh}\, [1]} = e^{{\rm i} \omega \zeta_s(r)} \widetilde {\delta A}_0^{{\rm nh}\, [1]}$ and $\widetilde {\delta A}_i^{{\rm nh}\, [0]} = e^{- {\rm i} \omega \zeta_s(r)} \delta A_i^{{\rm nh}\, [0]} $. The general solution is
\begin{align}
\delta A_0^{{\rm nh}\, [1]}(r) &= e^{{\rm i} \omega \zeta_s(r)} \left\{ \#_3 + \#_4 \int_{r_h}^r \frac{dy}{y^3[1+ 3 \mathcal C(y)]} \right. \nonumber \\
& \left.+ \int_{r_h}^r\frac{dy}{y^3[1+ 3 \mathcal C(y)]} \int_{r_h}^y \frac{\omega k_i}{z f(z)}(1 + \mathcal C(z)) \right. \nonumber \\
& \quad \widetilde{\delta A}_i^{{\rm nh}\,[0]}(z) dz\biggr\} ,\label{C0_nh1}
\end{align}
where $\#_{3,4}$ may be fixed by boundary conditions. In fact, one does not need to work out the integrals in Eq. \eqref{C0_nh1} given the following observation: $\delta A_0^{{\rm nh}\,[1]} \sim \partial_i \mathfrak J_i^{\rm nh}$, which stands for the non-hydro contribution to the conserved charge density \cite{Jain:2023obu}. Eventually, we just need to read off $\mathcal V_{si}$'s in Eq. \eqref{Cmu_boundary} from Eq. \eqref{Cinh_k0_sol},
\begin{align}
\mathcal V_{ri}(\omega) &= a(\omega) \mathfrak J_{ri}^{\rm nh}(\omega) + \frac{1}{2} {\rm coth}\left( \frac{\beta \omega}{2} \right)a(\omega) \mathfrak J_{ai}^{\rm nh}(\omega) \nonumber \\
&- \frac{1}{2} {\rm coth}\left( \frac{\beta \omega}{2} \right)a(-\omega) \mathfrak J_{ai}^{\rm nh}(\omega)\,, \nonumber \\
\mathcal V_{ai}(\omega) &= a(-\omega) \mathfrak J_{ai}^{\rm nh}(\omega)\,,\label{Ci_nh_non-normalizable}
\end{align}
where $a(\omega) = \phi(\omega)/\mathcal O(\omega)$.
Practically, we expand
Eq. \eqref{Cinh_k0_sol} around $\omega = \omega_1$ and pick the first non-hydro mode.

Using Eq. \eqref{Cmu_h_normalizable} and Eq. \eqref{Ci_nh_non-normalizable}, the quadratic bulk action \eqref{B+V_J} is $S_0^{(2)} = S_{\text{eff}}^{\rm h} + S_{\text{eff}}^{\rm nh}$ with
\begin{align}
S_{\text{eff}}^{\rm h} & = \int d^4x \left( \chi B_{a0} B_{r0} - \sigma_0 B_{ai} \partial_0 B_{ri} + {\rm i} \beta^{-1} \sigma B_{ai}^2 \right)\,,\nonumber \\
S_{\text{eff}}^{\rm nh} &= \int d^4x \biggr\{ - \frac{a_1}{2} \frac{\sigma_0}{\tau} \left( B_{ai} \partial_0 \mathfrak J_{ri}^{\rm nh} + \mathfrak J_{ai}^{\rm nh} \partial_0 B_{ri}\right. \nonumber \\
& -2 {\rm i} \beta^{-1} B_{ai} \mathfrak J_{ai}^{\rm nh}\left. \right) + 2a_1 \mathfrak J_{ai}^{\rm nh} (\partial_0 + \tau^{-1}) \mathfrak J_{ri}^{\rm nh} \nonumber \\
& + {\rm i} a_1 (\partial_0 + \tau^{-1}) \left[\cot\left( \frac{\beta}{2\tau}\right)  \right. \nonumber \\
& + (\partial_0 + \tau^{-1})\frac{\beta}{2\tau} \sin^{-2}\left(\frac{\beta}{2\tau} \right)\biggr] \mathfrak J_{ai}^{\rm nh} \mathfrak J_{ai}^{\rm nh}
 \biggr\}\,,\label{Seff_holo}
\end{align}
where $\beta = 1/T$, and the various coefficients are given by
\begin{align}
& \chi = \left( \int_{r_h}^{\infty} \frac{dy}{y^3 [1+3 \mathcal C(y)]} \right)^{-1}\,, \qquad \sigma_0 =  r_h \left[ 1 + \mathcal C(r_h) \right]\,,\nonumber \\
& \tau = - \frac{ \rm i}{\omega_1}, \qquad a_1 = {\rm i} a^\prime(\omega_1)\,.
\end{align}

The holographic action \eqref{Seff_holo} enables us to test, from first principles, the fate of chemical shift symmetry and the dynamical KMS symmetry in effective field theories that include non-hydrodynamic modes. From a purely EFT perspective, it remains unclear whether these symmetries extend to non-hydrodynamic fields and, if so, in what form.

Our results confirm that the chemical shift symmetry remains intact, in agreement with the assumption made in Ref.~\cite{Jain:2023obu}. More precisely, if chemical shift symmetry holds in the hydrodynamic sector, the inclusion of a non-hydrodynamic mode does not break it.

In contrast, the case of dynamical KMS symmetry is more subtle. We have verified that the action \eqref{Seff_holo} is invariant under the generalized KMS transformations
\begin{align}
\hat B_{r\mu}(-x)&  = B_{r\mu}(x)\,, \nonumber \\
\hat B_{a\mu}(-x)& =  B_{a\mu}(x) + i \beta \partial_0 B_{r\mu}(x)\,, \nonumber \\
\hat {\mathfrak J}_{ri}^{\rm nh}(-x) & =  e^{\frac{ {\rm i }\beta }{2 \tau}} \left(1 - {\rm i}\frac{\beta}{2} \partial_0^* \right) \left[\cos(\frac{\beta }{2 \tau} ) \right. \nonumber \\
& + \sin(\frac{\beta }{2 \tau} )\frac{\beta}{2} \partial_0^*  \biggr]\mathfrak J_{ri}^{\rm nh}(x) \nonumber \\
&-  \frac{1}{2} {\rm i} e^{\frac{ {\rm i }\beta }{2 \tau}} \left(1 - {\rm i}\frac{\beta}{2} \partial_0^* \right) \left[\sin(\frac{\beta }{2 \tau} ) \right. \nonumber \\
&- \cos(\frac{\beta }{2 \tau} )\frac{\beta}{2} \partial_0^* \biggr]\mathfrak J_{ai}^{\rm nh}(x) \,,\nonumber \\
\hat {\mathfrak J}_{ai}^{\rm nh}(-x)  & = e^{\frac{ {\rm i }\beta }{2 \tau}} \left(1 - {\rm i}\frac{\beta}{2} \partial_0^* \right) \left[\cos(\frac{\beta }{2 \tau} )\right. \nonumber \\
& + \sin(\frac{\beta }{2 \tau} )\frac{\beta}{2} \partial_0^* \biggr]\mathfrak J_{ai}^{\rm nh}(x) \nonumber \\
& - 2 {\rm i} e^{\frac{ {\rm i }\beta }{2 \tau}}\left(1 - {\rm i}\frac{\beta}{2} \partial_0^* \right) \left[\sin(\frac{\beta }{2 \tau} ) \right.\nonumber \\
&- \cos(\frac{\beta }{2 \tau} )\frac{\beta}{2} \partial_0^* \biggr] \mathfrak J_{ri}^{\rm nh}(x)\,, \label{KMS_B_Jnh}
\end{align}
where $\partial_0^* = \partial_0 + \tau^{-1}$. The classical statistical limit of more general KMS condition just found can be obtained by taking the limit $\partial_0 \to 0$, yielding to results consistent with those in \cite{Liu:2018kfw,Jain:2023obu}. Then, the action in Eq.~\eqref{Seff_holo} is the same as that in Eq.~(8) once identifying $\mathfrak J_i^{\rm nh}$ as $V_i$ and choosing the same hydrodynamic frame. At this point, the universality and fundamental origin of this generalized KMS transformation involving non-hydrodynamic degrees of freedom remain partially unknown.  In the near future, it would be interesting to investigate this aspect further in a more general setup involving non-hydrodynamic modes.

\subsection{Comments on the holographic derivation of boundary EFT with a non-hydro mode}\label{compi}
The decomposition employed in Eq.~\eqref{Cmu_solution} highlights a major distinction from Ref.~\cite{Liu:2024tqe}, where the non-hydrodynamic mode is introduced through Dirichlet boundary conditions applied at a stretched horizon. As a result, that method requires an extra ad-hoc step to map near-horizon data to boundary variables, to match the expressions in Ref.~\cite{Jain:2023obu}. In contrast, our approach directly relates the non-hydro variable to the spatial boundary flux, making the physical interpretation more transparent. Furthermore, the method in Ref.~\cite{Liu:2024tqe} appears to incorrectly capture the symmetry structure of the dual EFT, particularly with regard to the chemical shift symmetry and the breakdown of the canonical KMS symmetry.

\subsection{Generalized KMS symmetry for non-hydro modes}
Our holographic result, Eq.~\eqref{Seff_holo}, allows us to address a fundamental question: what is the fate of chemical shift symmetry and dynamical KMS symmetry, two cornerstone symmetries in the formulation of {\it hydrodynamic} EFTs~\cite{Liu:2018kfw}, in the presence of non-hydrodynamic modes? From a purely EFT perspective, it remains unclear whether these symmetries extend to non-hydrodynamic fields and, if so, in what form. Previous works, such as Ref.~\cite{Jain:2023obu}, assumed their applicability without providing physical justification or explicit derivation.

Our results confirm that chemical shift symmetry remains intact, in agreement with the assumption made in Ref.~\cite{Jain:2023obu}. More precisely, if chemical shift symmetry holds in the hydrodynamic sector, the inclusion of a non-hydrodynamic mode does not break it.

In contrast, the case of dynamical KMS symmetry is more subtle. Our analysis reveals that the effective action involving non-hydrodynamic modes does not respect the canonical form of KMS symmetry, contrary to the assumption in Ref.~\cite{Jain:2023obu}. Instead, the action is invariant under a generalized version of the KMS symmetry, defined by the transformations in Eqs.~\eqref{KMS_B_Jnh}.

These findings provide strong evidence that the dynamical KMS symmetry acts fundamentally differently on hydrodynamic versus non-hydrodynamic variables. This distinction is conceptually justified by the fluctuation-dissipation relation,
\begin{equation}
G_S(\omega, \vec{k}) = \coth \left( \frac{\beta \omega}{2} \right) \mathrm{Im} \left[ G_R(\omega, \vec{k}) \right]\,,
\end{equation}
which is guaranteed by KMS symmetry and remains valid for arbitrary $\omega$, including $\omega= - {\rm i}/\tau$ as relevant for non-hydrodynamic modes.

The classical statistical limit of the generalized KMS symmetry can be recovered by taking $\partial_0 \to 0$, yielding results consistent with those of Refs.~\cite{Liu:2018kfw, Jain:2023obu}. In this limit, the action, Eq.~\eqref{Seff_holo}, exactly reduces to that in Eq.~\eqref{MCeft}, once the identification $\mathfrak J_i^{\rm nh} \equiv V_i$ is made and the same hydrodynamic frame is adopted.
\section{Outlook}
In this work, we have constructed a simple holographic dual for the relativistic Maxwell-Cattaneo model. The equivalence between the gravitational framework and the dual effective description is verified combining numerical and analytical methods. These analyses not only validate the structure of the MC model in the dual field theory but also provide analytical approximated formulae  in terms of holographic data for all the parameters appearing therein.

Importantly, the way in which the holographic model realizes Maxwell-Cattaneo (MC) dynamics is remarkably natural. The MC model can be viewed as a ultraviolet (UV) completion of the standard diffusion equation, modifying its behavior at large frequencies and wavevectors. In the holographic framework, this role is played by the $(F^2)^2$ term in the bulk action, which introduces UV corrections through higher-derivative terms. These corrections modify the high-energy dynamics of the original Maxwell theory in the bulk and carry the same physical meaning as the relaxation time $\tau$ in the dual field theory. We also note a striking analogy with the Israel-Stewart formalism, which similarly requires Gauss-Bonnet corrections in the bulk to capture causal relativistic hydrodynamics \cite{Grozdanov:2018fic}.

We anticipate that our holographic model could have direct applications in condensed matter systems where strong electron-electron interactions give rise to hydrodynamic charge dynamics (see \cite{Varnavides2023} for a recent review). In fact, the relaxation time $\tau$, which, consistent with the Maxwell-Cattaneo formalism, modifies the diffusive behavior of charge density fluctuations, has already been experimentally measured in an ultracold lithium-6 gas confined to a two-dimensional optical lattice \cite{doi:10.1126/science.aat4134}.

Having a simple holographic dual of the MC model at hand extends the applicability of holographic methods to shorter time-scales, with possible important implications to the study of strongly coupled fluids such as the quark-gluon plasma \cite{PhysRevLett.130.212303}, and in particular their non-equilibrium physics as well. For example, it allows the study of the interactions between the hydrodynamic fields and the non-hydro variables, with potential important physical implications (see for example \cite{PhysRevLett.132.131602}).

\section*{Acknowledgments}  We would like to thank Yi Yin, Yan Liu, Andrea Amoretti and Xin-Meng Wu for useful discussions. YA and MB acknowledge the support of the Shanghai Municipal Science and Technology Major Project (Grant No.2019SHZDZX01). MB acknowledges the support of the sponsorship from the Yangyang Development Fund. YB and XS were supported by the National Natural Science Foundation of China
(NSFC) under the grant No. 12375044.


\begin{thebibliography}{53}%
\makeatletter
\providecommand \@ifxundefined [1]{%
 \@ifx{#1\undefined}
}%
\providecommand \@ifnum [1]{%
 \ifnum #1\expandafter \@firstoftwo
 \else \expandafter \@secondoftwo
 \fi
}%
\providecommand \@ifx [1]{%
 \ifx #1\expandafter \@firstoftwo
 \else \expandafter \@secondoftwo
 \fi
}%
\providecommand \natexlab [1]{#1}%
\providecommand \enquote  [1]{``#1''}%
\providecommand \bibnamefont  [1]{#1}%
\providecommand \bibfnamefont [1]{#1}%
\providecommand \citenamefont [1]{#1}%
\providecommand \href@noop [0]{\@secondoftwo}%
\providecommand \href [0]{\begingroup \@sanitize@url \@href}%
\providecommand \@href[1]{\@@startlink{#1}\@@href}%
\providecommand \@@href[1]{\endgroup#1\@@endlink}%
\providecommand \@sanitize@url [0]{\catcode `\\12\catcode `\$12\catcode `\&12\catcode `\#12\catcode `\^12\catcode `\_12\catcode `\%12\relax}%
\providecommand \@@startlink[1]{}%
\providecommand \@@endlink[0]{}%
\providecommand \url  [0]{\begingroup\@sanitize@url \@url }%
\providecommand \@url [1]{\endgroup\@href {#1}{\urlprefix }}%
\providecommand \urlprefix  [0]{URL }%
\providecommand \Eprint [0]{\href }%
\providecommand \doibase [0]{http://dx.doi.org/}%
\providecommand \selectlanguage [0]{\@gobble}%
\providecommand \bibinfo  [0]{\@secondoftwo}%
\providecommand \bibfield  [0]{\@secondoftwo}%
\providecommand \translation [1]{[#1]}%
\providecommand \BibitemOpen [0]{}%
\providecommand \bibitemStop [0]{}%
\providecommand \bibitemNoStop [0]{.\EOS\space}%
\providecommand \EOS [0]{\spacefactor3000\relax}%
\providecommand \BibitemShut  [1]{\csname bibitem#1\endcsname}%
\let\auto@bib@innerbib\@empty
\bibitem [{\citenamefont {Narasimhan}(2009)}]{10.1063/1.3177228}%
  \BibitemOpen
  \bibfield  {author} {\bibinfo {author} {\bibfnamefont {T.~N.}\ \bibnamefont {Narasimhan}},\ }\href {\doibase 10.1063/1.3177228} {\bibfield  {journal} {\bibinfo  {journal} {Physics Today}\ }\textbf {\bibinfo {volume} {62}},\ \bibinfo {pages} {48} (\bibinfo {year} {2009})}\BibitemShut {NoStop}%
\bibitem [{\citenamefont {and}(1855)}]{Fick01071855}%
  \BibitemOpen
  \bibfield  {author} {\bibinfo {author} {\bibfnamefont {A.~F.}\ \bibnamefont {and}},\ }\href {\doibase 10.1080/14786445508641925} {\bibfield  {journal} {\bibinfo  {journal} {The London, Edinburgh, and Dublin Philosophical Magazine and Journal of Science}\ }\textbf {\bibinfo {volume} {10}},\ \bibinfo {pages} {30} (\bibinfo {year} {1855})}\BibitemShut {NoStop}%
\bibitem [{\citenamefont {Joseph}\ and\ \citenamefont {Preziosi}(1989)}]{RevModPhys.61.41}%
  \BibitemOpen
  \bibfield  {author} {\bibinfo {author} {\bibfnamefont {D.~D.}\ \bibnamefont {Joseph}}\ and\ \bibinfo {author} {\bibfnamefont {L.}~\bibnamefont {Preziosi}},\ }\href {\doibase 10.1103/RevModPhys.61.41} {\bibfield  {journal} {\bibinfo  {journal} {Rev. Mod. Phys.}\ }\textbf {\bibinfo {volume} {61}},\ \bibinfo {pages} {41} (\bibinfo {year} {1989})}\BibitemShut {NoStop}%
\bibitem [{\citenamefont {Jou}\ \emph {et~al.}(1988)\citenamefont {Jou}, \citenamefont {Casas-Vazquez},\ and\ \citenamefont {Lebon}}]{Jou_1988}%
  \BibitemOpen
  \bibfield  {author} {\bibinfo {author} {\bibfnamefont {D.}~\bibnamefont {Jou}}, \bibinfo {author} {\bibfnamefont {J.}~\bibnamefont {Casas-Vazquez}}, \ and\ \bibinfo {author} {\bibfnamefont {G.}~\bibnamefont {Lebon}},\ }\href {\doibase 10.1088/0034-4885/51/8/002} {\bibfield  {journal} {\bibinfo  {journal} {Reports on Progress in Physics}\ }\textbf {\bibinfo {volume} {51}},\ \bibinfo {pages} {1105} (\bibinfo {year} {1988})}\BibitemShut {NoStop}%
\bibitem [{\citenamefont {Chandrasekharaiah}(1986)}]{10.1115/1.3143705}%
  \BibitemOpen
  \bibfield  {author} {\bibinfo {author} {\bibfnamefont {D.~S.}\ \bibnamefont {Chandrasekharaiah}},\ }\href {\doibase 10.1115/1.3143705} {\bibfield  {journal} {\bibinfo  {journal} {Applied Mechanics Reviews}\ }\textbf {\bibinfo {volume} {39}},\ \bibinfo {pages} {355} (\bibinfo {year} {1986})}\BibitemShut {NoStop}%
\bibitem [{\citenamefont {Cattaneo}(1948)}]{1573950400546202112}%
  \BibitemOpen
  \bibfield  {author} {\bibinfo {author} {\bibfnamefont {C.}~\bibnamefont {Cattaneo}},\ }\href {https://cir.nii.ac.jp/crid/1573950400546202112} {\bibfield  {journal} {\bibinfo  {journal} {Atti Sem. Mat. Fis. Univ. Modena}\ }\textbf {\bibinfo {volume} {3}},\ \bibinfo {pages} {83} (\bibinfo {year} {1948})}\BibitemShut {NoStop}%
\bibitem [{\citenamefont {Maxwell}(1867)}]{doi:10.1098/rstl.1867.0004}%
  \BibitemOpen
  \bibfield  {author} {\bibinfo {author} {\bibfnamefont {J.~C.}\ \bibnamefont {Maxwell}},\ }\href {\doibase 10.1098/rstl.1867.0004} {\bibfield  {journal} {\bibinfo  {journal} {Philosophical Transactions of the Royal Society of London}\ }\textbf {\bibinfo {volume} {157}},\ \bibinfo {pages} {49} (\bibinfo {year} {1867})}\BibitemShut {NoStop}%
\bibitem [{\citenamefont {Vernotte}(1958)}]{1571980075311404160}%
  \BibitemOpen
  \bibfield  {author} {\bibinfo {author} {\bibfnamefont {P.}~\bibnamefont {Vernotte}},\ }\href {https://cir.nii.ac.jp/crid/1571980075311404160} {\bibfield  {journal} {\bibinfo  {journal} {Compt. Rendu}\ }\textbf {\bibinfo {volume} {246}},\ \bibinfo {pages} {3154} (\bibinfo {year} {1958})}\BibitemShut {NoStop}%
\bibitem [{\citenamefont {Heaviside}(1894)}]{heaviside1894electrical}%
  \BibitemOpen
  \bibfield  {author} {\bibinfo {author} {\bibfnamefont {O.}~\bibnamefont {Heaviside}},\ }\href {https://books.google.com.tw/books?id=bywPAAAAIAAJ} {\emph {\bibinfo {title} {Electrical Papers}}},\ \bibinfo {series} {Electrical Papers}\ No.\ \bibinfo {number} {v. 2}\ (\bibinfo  {publisher} {Macmillan and Company},\ \bibinfo {year} {1894})\BibitemShut {NoStop}%
\bibitem [{\citenamefont {Baggioli}\ \emph {et~al.}(2020{\natexlab{a}})\citenamefont {Baggioli}, \citenamefont {Vasin}, \citenamefont {Brazhkin},\ and\ \citenamefont {Trachenko}}]{BAGGIOLI20201}%
  \BibitemOpen
  \bibfield  {author} {\bibinfo {author} {\bibfnamefont {M.}~\bibnamefont {Baggioli}}, \bibinfo {author} {\bibfnamefont {M.}~\bibnamefont {Vasin}}, \bibinfo {author} {\bibfnamefont {V.}~\bibnamefont {Brazhkin}}, \ and\ \bibinfo {author} {\bibfnamefont {K.}~\bibnamefont {Trachenko}},\ }\href {\doibase https://doi.org/10.1016/j.physrep.2020.04.002} {\bibfield  {journal} {\bibinfo  {journal} {Physics Reports}\ }\textbf {\bibinfo {volume} {865}},\ \bibinfo {pages} {1} (\bibinfo {year} {2020}{\natexlab{a}})},\ \bibinfo {note} {gapped momentum states}\BibitemShut {NoStop}%
\bibitem [{\citenamefont {Nosenko}\ \emph {et~al.}(2006)\citenamefont {Nosenko}, \citenamefont {Goree},\ and\ \citenamefont {Piel}}]{PhysRevLett.97.115001}%
  \BibitemOpen
  \bibfield  {author} {\bibinfo {author} {\bibfnamefont {V.}~\bibnamefont {Nosenko}}, \bibinfo {author} {\bibfnamefont {J.}~\bibnamefont {Goree}}, \ and\ \bibinfo {author} {\bibfnamefont {A.}~\bibnamefont {Piel}},\ }\href {\doibase 10.1103/PhysRevLett.97.115001} {\bibfield  {journal} {\bibinfo  {journal} {Phys. Rev. Lett.}\ }\textbf {\bibinfo {volume} {97}},\ \bibinfo {pages} {115001} (\bibinfo {year} {2006})}\BibitemShut {NoStop}%
\bibitem [{\citenamefont {Jiang}\ \emph {et~al.}(2025)\citenamefont {Jiang}, \citenamefont {Zheng}, \citenamefont {Chen}, \citenamefont {Baggioli},\ and\ \citenamefont {Zhang}}]{Jiang2025}%
  \BibitemOpen
  \bibfield  {author} {\bibinfo {author} {\bibfnamefont {C.}~\bibnamefont {Jiang}}, \bibinfo {author} {\bibfnamefont {Z.}~\bibnamefont {Zheng}}, \bibinfo {author} {\bibfnamefont {Y.}~\bibnamefont {Chen}}, \bibinfo {author} {\bibfnamefont {M.}~\bibnamefont {Baggioli}}, \ and\ \bibinfo {author} {\bibfnamefont {J.}~\bibnamefont {Zhang}},\ }\href {\doibase 10.1038/s42005-025-02008-1} {\bibfield  {journal} {\bibinfo  {journal} {Communications Physics}\ }\textbf {\bibinfo {volume} {8}},\ \bibinfo {pages} {82} (\bibinfo {year} {2025})}\BibitemShut {NoStop}%
\bibitem [{\citenamefont {Bai}\ \emph {et~al.}(2025)\citenamefont {Bai}, \citenamefont {Keim},\ and\ \citenamefont {Baggioli}}]{bai2025tracking}%
  \BibitemOpen
  \bibfield  {author} {\bibinfo {author} {\bibfnamefont {J.}~\bibnamefont {Bai}}, \bibinfo {author} {\bibfnamefont {P.}~\bibnamefont {Keim}}, \ and\ \bibinfo {author} {\bibfnamefont {M.}~\bibnamefont {Baggioli}},\ }\href@noop {} {\bibfield  {journal} {\bibinfo  {journal} {arXiv preprint arXiv:2505.16678}\ } (\bibinfo {year} {2025})}\BibitemShut {NoStop}%
\bibitem [{\citenamefont {Chester}(1963)}]{PhysRev.131.2013}%
  \BibitemOpen
  \bibfield  {author} {\bibinfo {author} {\bibfnamefont {M.}~\bibnamefont {Chester}},\ }\href {\doibase 10.1103/PhysRev.131.2013} {\bibfield  {journal} {\bibinfo  {journal} {Phys. Rev.}\ }\textbf {\bibinfo {volume} {131}},\ \bibinfo {pages} {2013} (\bibinfo {year} {1963})}\BibitemShut {NoStop}%
\bibitem [{\citenamefont {Brown}\ \emph {et~al.}(2019)\citenamefont {Brown}, \citenamefont {Mitra}, \citenamefont {Guardado-Sanchez}, \citenamefont {Nourafkan}, \citenamefont {Reymbaut}, \citenamefont {Hebert}, \citenamefont {Bergeron}, \citenamefont {Tremblay}, \citenamefont {Kokalj}, \citenamefont {Huse}, \citenamefont {Schaub},\ and\ \citenamefont {Bakr}}]{doi:10.1126/science.aat4134}%
  \BibitemOpen
  \bibfield  {author} {\bibinfo {author} {\bibfnamefont {P.~T.}\ \bibnamefont {Brown}}, \bibinfo {author} {\bibfnamefont {D.}~\bibnamefont {Mitra}}, \bibinfo {author} {\bibfnamefont {E.}~\bibnamefont {Guardado-Sanchez}}, \bibinfo {author} {\bibfnamefont {R.}~\bibnamefont {Nourafkan}}, \bibinfo {author} {\bibfnamefont {A.}~\bibnamefont {Reymbaut}}, \bibinfo {author} {\bibfnamefont {C.-D.}\ \bibnamefont {Hebert}}, \bibinfo {author} {\bibfnamefont {S.}~\bibnamefont {Bergeron}}, \bibinfo {author} {\bibfnamefont {A.-M.~S.}\ \bibnamefont {Tremblay}}, \bibinfo {author} {\bibfnamefont {J.}~\bibnamefont {Kokalj}}, \bibinfo {author} {\bibfnamefont {D.~A.}\ \bibnamefont {Huse}}, \bibinfo {author} {\bibfnamefont {P.}~\bibnamefont {Schaub}}, \ and\ \bibinfo {author} {\bibfnamefont {W.~S.}\ \bibnamefont {Bakr}},\ }\href {\doibase 10.1126/science.aat4134} {\bibfield  {journal} {\bibinfo  {journal} {Science}\ }\textbf {\bibinfo {volume} {363}},\ \bibinfo {pages} {379} (\bibinfo {year} {2019})}\BibitemShut {NoStop}%
\bibitem [{\citenamefont {Romatschke}\ and\ \citenamefont {Romatschke}(2019)}]{Romatschke:2017ejr}%
  \BibitemOpen
  \bibfield  {author} {\bibinfo {author} {\bibfnamefont {P.}~\bibnamefont {Romatschke}}\ and\ \bibinfo {author} {\bibfnamefont {U.}~\bibnamefont {Romatschke}},\ }\href {\doibase 10.1017/9781108651998} {\emph {\bibinfo {title} {{Relativistic Fluid Dynamics In and Out of Equilibrium}}}},\ Cambridge Monographs on Mathematical Physics\ (\bibinfo  {publisher} {Cambridge University Press},\ \bibinfo {year} {2019})\ \Eprint {http://arxiv.org/abs/1712.05815} {arXiv:1712.05815 [nucl-th]} \BibitemShut {NoStop}%
\bibitem [{\citenamefont {Gale}\ \emph {et~al.}(2013)\citenamefont {Gale}, \citenamefont {Jeon},\ and\ \citenamefont {Schenke}}]{Gale:2013da}%
  \BibitemOpen
  \bibfield  {author} {\bibinfo {author} {\bibfnamefont {C.}~\bibnamefont {Gale}}, \bibinfo {author} {\bibfnamefont {S.}~\bibnamefont {Jeon}}, \ and\ \bibinfo {author} {\bibfnamefont {B.}~\bibnamefont {Schenke}},\ }\href {\doibase 10.1142/S0217751X13400113} {\bibfield  {journal} {\bibinfo  {journal} {Int. J. Mod. Phys. A}\ }\textbf {\bibinfo {volume} {28}},\ \bibinfo {pages} {1340011} (\bibinfo {year} {2013})},\ \Eprint {http://arxiv.org/abs/1301.5893} {arXiv:1301.5893 [nucl-th]} \BibitemShut {NoStop}%
\bibitem [{\citenamefont {Kovtun}(2012)}]{Kovtun:2012rj}%
  \BibitemOpen
  \bibfield  {author} {\bibinfo {author} {\bibfnamefont {P.}~\bibnamefont {Kovtun}},\ }\href {\doibase 10.1088/1751-8113/45/47/473001} {\bibfield  {journal} {\bibinfo  {journal} {J. Phys. A}\ }\textbf {\bibinfo {volume} {45}},\ \bibinfo {pages} {473001} (\bibinfo {year} {2012})},\ \Eprint {http://arxiv.org/abs/1205.5040} {arXiv:1205.5040 [hep-th]} \BibitemShut {NoStop}%
\bibitem [{\citenamefont {Israel}\ and\ \citenamefont {Stewart}(1979)}]{Israel:1979wp}%
  \BibitemOpen
  \bibfield  {author} {\bibinfo {author} {\bibfnamefont {W.}~\bibnamefont {Israel}}\ and\ \bibinfo {author} {\bibfnamefont {J.~M.}\ \bibnamefont {Stewart}},\ }\href {\doibase 10.1016/0003-4916(79)90130-1} {\bibfield  {journal} {\bibinfo  {journal} {Annals Phys.}\ }\textbf {\bibinfo {volume} {118}},\ \bibinfo {pages} {341} (\bibinfo {year} {1979})}\BibitemShut {NoStop}%
\bibitem [{\citenamefont {Jain}\ and\ \citenamefont {Kovtun}(2024)}]{Jain:2023obu}%
  \BibitemOpen
  \bibfield  {author} {\bibinfo {author} {\bibfnamefont {A.}~\bibnamefont {Jain}}\ and\ \bibinfo {author} {\bibfnamefont {P.}~\bibnamefont {Kovtun}},\ }\href {\doibase 10.1007/JHEP01(2024)162} {\bibfield  {journal} {\bibinfo  {journal} {JHEP}\ }\textbf {\bibinfo {volume} {01}},\ \bibinfo {pages} {162} (\bibinfo {year} {2024})},\ \Eprint {http://arxiv.org/abs/2309.00511} {arXiv:2309.00511 [hep-th]} \BibitemShut {NoStop}%
\bibitem [{\citenamefont {Hongo}\ \emph {et~al.}(2024)\citenamefont {Hongo}, \citenamefont {Sogabe}, \citenamefont {Stephanov},\ and\ \citenamefont {Yee}}]{Hongo:2024brb}%
  \BibitemOpen
  \bibfield  {author} {\bibinfo {author} {\bibfnamefont {M.}~\bibnamefont {Hongo}}, \bibinfo {author} {\bibfnamefont {N.}~\bibnamefont {Sogabe}}, \bibinfo {author} {\bibfnamefont {M.~A.}\ \bibnamefont {Stephanov}}, \ and\ \bibinfo {author} {\bibfnamefont {H.-U.}\ \bibnamefont {Yee}},\ }\href@noop {} {\  (\bibinfo {year} {2024})},\ \Eprint {http://arxiv.org/abs/2411.08016} {arXiv:2411.08016 [hep-th]} \BibitemShut {NoStop}%
\bibitem [{\citenamefont {Baggioli}\ \emph {et~al.}(2020{\natexlab{b}})\citenamefont {Baggioli}, \citenamefont {Vasin}, \citenamefont {Brazhkin},\ and\ \citenamefont {Trachenko}}]{Baggioli:2020whu}%
  \BibitemOpen
  \bibfield  {author} {\bibinfo {author} {\bibfnamefont {M.}~\bibnamefont {Baggioli}}, \bibinfo {author} {\bibfnamefont {M.}~\bibnamefont {Vasin}}, \bibinfo {author} {\bibfnamefont {V.~V.}\ \bibnamefont {Brazhkin}}, \ and\ \bibinfo {author} {\bibfnamefont {K.}~\bibnamefont {Trachenko}},\ }\href {\doibase 10.1103/PhysRevD.102.025012} {\bibfield  {journal} {\bibinfo  {journal} {Phys. Rev. D}\ }\textbf {\bibinfo {volume} {102}},\ \bibinfo {pages} {025012} (\bibinfo {year} {2020}{\natexlab{b}})},\ \Eprint {http://arxiv.org/abs/2004.13613} {arXiv:2004.13613 [hep-th]} \BibitemShut {NoStop}%
\bibitem [{\citenamefont {Martinoia}(2024)}]{Martinoia:2024cbw}%
  \BibitemOpen
  \bibfield  {author} {\bibinfo {author} {\bibfnamefont {L.}~\bibnamefont {Martinoia}},\ }\emph {\bibinfo {title} {{Developments in quasihydrodynamics}}},\ \href {\doibase 10.15167/martinoia-luca_phd2024-03-01} {Ph.D. thesis},\ \bibinfo  {school} {Genoa U.} (\bibinfo {year} {2024}),\ \Eprint {http://arxiv.org/abs/2403.14254} {arXiv:2403.14254 [hep-th]} \BibitemShut {NoStop}%
\bibitem [{\citenamefont {Stephanov}\ and\ \citenamefont {Yin}(2018{\natexlab{a}})}]{PhysRevD.98.036006}%
  \BibitemOpen
  \bibfield  {author} {\bibinfo {author} {\bibfnamefont {M.}~\bibnamefont {Stephanov}}\ and\ \bibinfo {author} {\bibfnamefont {Y.}~\bibnamefont {Yin}},\ }\href {\doibase 10.1103/PhysRevD.98.036006} {\bibfield  {journal} {\bibinfo  {journal} {Phys. Rev. D}\ }\textbf {\bibinfo {volume} {98}},\ \bibinfo {pages} {036006} (\bibinfo {year} {2018}{\natexlab{a}})}\BibitemShut {NoStop}%
\bibitem [{\citenamefont {Pradeep}\ \emph {et~al.}(2022)\citenamefont {Pradeep}, \citenamefont {Rajagopal}, \citenamefont {Stephanov},\ and\ \citenamefont {Yin}}]{PhysRevD.106.036017}%
  \BibitemOpen
  \bibfield  {author} {\bibinfo {author} {\bibfnamefont {M.}~\bibnamefont {Pradeep}}, \bibinfo {author} {\bibfnamefont {K.}~\bibnamefont {Rajagopal}}, \bibinfo {author} {\bibfnamefont {M.}~\bibnamefont {Stephanov}}, \ and\ \bibinfo {author} {\bibfnamefont {Y.}~\bibnamefont {Yin}},\ }\href {\doibase 10.1103/PhysRevD.106.036017} {\bibfield  {journal} {\bibinfo  {journal} {Phys. Rev. D}\ }\textbf {\bibinfo {volume} {106}},\ \bibinfo {pages} {036017} (\bibinfo {year} {2022})}\BibitemShut {NoStop}%
\bibitem [{\citenamefont {Ke}\ and\ \citenamefont {Yin}(2023)}]{PhysRevLett.130.212303}%
  \BibitemOpen
  \bibfield  {author} {\bibinfo {author} {\bibfnamefont {W.}~\bibnamefont {Ke}}\ and\ \bibinfo {author} {\bibfnamefont {Y.}~\bibnamefont {Yin}},\ }\href {\doibase 10.1103/PhysRevLett.130.212303} {\bibfield  {journal} {\bibinfo  {journal} {Phys. Rev. Lett.}\ }\textbf {\bibinfo {volume} {130}},\ \bibinfo {pages} {212303} (\bibinfo {year} {2023})}\BibitemShut {NoStop}%
\bibitem [{\citenamefont {Chandrasekharaiah}(1998)}]{10.1115/1.3098984}%
  \BibitemOpen
  \bibfield  {author} {\bibinfo {author} {\bibfnamefont {D.~S.}\ \bibnamefont {Chandrasekharaiah}},\ }\href {\doibase 10.1115/1.3098984} {\bibfield  {journal} {\bibinfo  {journal} {Applied Mechanics Reviews}\ }\textbf {\bibinfo {volume} {51}},\ \bibinfo {pages} {705} (\bibinfo {year} {1998})}\BibitemShut {NoStop}%
\bibitem [{\citenamefont {Stephanov}\ and\ \citenamefont {Yin}(2018{\natexlab{b}})}]{Stephanov:2017ghc}%
  \BibitemOpen
  \bibfield  {author} {\bibinfo {author} {\bibfnamefont {M.}~\bibnamefont {Stephanov}}\ and\ \bibinfo {author} {\bibfnamefont {Y.}~\bibnamefont {Yin}},\ }\href {\doibase 10.1103/PhysRevD.98.036006} {\bibfield  {journal} {\bibinfo  {journal} {Phys. Rev. D}\ }\textbf {\bibinfo {volume} {98}},\ \bibinfo {pages} {036006} (\bibinfo {year} {2018}{\natexlab{b}})},\ \Eprint {http://arxiv.org/abs/1712.10305} {arXiv:1712.10305 [nucl-th]} \BibitemShut {NoStop}%
\bibitem [{\citenamefont {Donos}\ and\ \citenamefont {Pantelidou}(2022)}]{Donos:2022xfd}%
  \BibitemOpen
  \bibfield  {author} {\bibinfo {author} {\bibfnamefont {A.}~\bibnamefont {Donos}}\ and\ \bibinfo {author} {\bibfnamefont {C.}~\bibnamefont {Pantelidou}},\ }\href {\doibase 10.1007/JHEP08(2022)246} {\bibfield  {journal} {\bibinfo  {journal} {JHEP}\ }\textbf {\bibinfo {volume} {08}},\ \bibinfo {pages} {246} (\bibinfo {year} {2022})},\ \Eprint {http://arxiv.org/abs/2205.06294} {arXiv:2205.06294 [hep-th]} \BibitemShut {NoStop}%
\bibitem [{\citenamefont {Donos}\ and\ \citenamefont {Kailidis}(2022)}]{Donos:2022qao}%
  \BibitemOpen
  \bibfield  {author} {\bibinfo {author} {\bibfnamefont {A.}~\bibnamefont {Donos}}\ and\ \bibinfo {author} {\bibfnamefont {P.}~\bibnamefont {Kailidis}},\ }\href {\doibase 10.1007/JHEP12(2022)028} {\bibfield  {journal} {\bibinfo  {journal} {JHEP}\ }\textbf {\bibinfo {volume} {12}},\ \bibinfo {pages} {028} (\bibinfo {year} {2022})},\ \bibinfo {note} {[Erratum: JHEP 07, 232 (2023)]},\ \Eprint {http://arxiv.org/abs/2210.06513} {arXiv:2210.06513 [hep-th]} \BibitemShut {NoStop}%
\bibitem [{\citenamefont {Donos}\ and\ \citenamefont {Kailidis}(2024)}]{Donos:2023ibv}%
  \BibitemOpen
  \bibfield  {author} {\bibinfo {author} {\bibfnamefont {A.}~\bibnamefont {Donos}}\ and\ \bibinfo {author} {\bibfnamefont {P.}~\bibnamefont {Kailidis}},\ }\href {\doibase 10.1007/JHEP01(2024)110} {\bibfield  {journal} {\bibinfo  {journal} {JHEP}\ }\textbf {\bibinfo {volume} {01}},\ \bibinfo {pages} {110} (\bibinfo {year} {2024})},\ \Eprint {http://arxiv.org/abs/2304.06008} {arXiv:2304.06008 [hep-th]} \BibitemShut {NoStop}%
\bibitem [{\citenamefont {Stephanov}\ and\ \citenamefont {Yin}(2017)}]{Stephanov:2017wlw}%
  \BibitemOpen
  \bibfield  {author} {\bibinfo {author} {\bibfnamefont {M.}~\bibnamefont {Stephanov}}\ and\ \bibinfo {author} {\bibfnamefont {Y.}~\bibnamefont {Yin}},\ }\href {\doibase 10.1016/j.nuclphysa.2017.06.051} {\bibfield  {journal} {\bibinfo  {journal} {Nucl. Phys. A}\ }\textbf {\bibinfo {volume} {967}},\ \bibinfo {pages} {876} (\bibinfo {year} {2017})},\ \Eprint {http://arxiv.org/abs/1704.07396} {arXiv:1704.07396 [nucl-th]} \BibitemShut {NoStop}%
\bibitem [{\citenamefont {Grozdanov}\ \emph {et~al.}(2019)\citenamefont {Grozdanov}, \citenamefont {Lucas},\ and\ \citenamefont {Poovuttikul}}]{Grozdanov:2018fic}%
  \BibitemOpen
  \bibfield  {author} {\bibinfo {author} {\bibfnamefont {S.}~\bibnamefont {Grozdanov}}, \bibinfo {author} {\bibfnamefont {A.}~\bibnamefont {Lucas}}, \ and\ \bibinfo {author} {\bibfnamefont {N.}~\bibnamefont {Poovuttikul}},\ }\href {\doibase 10.1103/PhysRevD.99.086012} {\bibfield  {journal} {\bibinfo  {journal} {Phys. Rev. D}\ }\textbf {\bibinfo {volume} {99}},\ \bibinfo {pages} {086012} (\bibinfo {year} {2019})},\ \Eprint {http://arxiv.org/abs/1810.10016} {arXiv:1810.10016 [hep-th]} \BibitemShut {NoStop}%
\bibitem [{\citenamefont {Policastro}\ \emph {et~al.}(2002)\citenamefont {Policastro}, \citenamefont {Son},\ and\ \citenamefont {Starinets}}]{GiuseppePolicastro_2002}%
  \BibitemOpen
  \bibfield  {author} {\bibinfo {author} {\bibfnamefont {G.}~\bibnamefont {Policastro}}, \bibinfo {author} {\bibfnamefont {D.~T.}\ \bibnamefont {Son}}, \ and\ \bibinfo {author} {\bibfnamefont {A.~O.}\ \bibnamefont {Starinets}},\ }\href {\doibase 10.1088/1126-6708/2002/09/043} {\bibfield  {journal} {\bibinfo  {journal} {Journal of High Energy Physics}\ }\textbf {\bibinfo {volume} {2002}},\ \bibinfo {pages} {043} (\bibinfo {year} {2002})}\BibitemShut {NoStop}%
\bibitem [{\citenamefont {Casalderrey-Solana}\ \emph {et~al.}(2014)\citenamefont {Casalderrey-Solana}, \citenamefont {Liu}, \citenamefont {Mateos}, \citenamefont {Rajagopal},\ and\ \citenamefont {Achim~Wiedemann}}]{Casalderrey-Solana:2011dxg}%
  \BibitemOpen
  \bibfield  {author} {\bibinfo {author} {\bibfnamefont {J.}~\bibnamefont {Casalderrey-Solana}}, \bibinfo {author} {\bibfnamefont {H.}~\bibnamefont {Liu}}, \bibinfo {author} {\bibfnamefont {D.}~\bibnamefont {Mateos}}, \bibinfo {author} {\bibfnamefont {K.}~\bibnamefont {Rajagopal}}, \ and\ \bibinfo {author} {\bibfnamefont {U.}~\bibnamefont {Achim~Wiedemann}},\ }\href {\doibase 10.1017/9781009403504} {\emph {\bibinfo {title} {{Gauge/String Duality, Hot QCD and Heavy Ion Collisions}}}}\ (\bibinfo  {publisher} {Cambridge University Press},\ \bibinfo {year} {2014})\ \Eprint {http://arxiv.org/abs/1101.0618} {arXiv:1101.0618 [hep-th]} \BibitemShut {NoStop}%
\bibitem [{\citenamefont {Baggioli}\ \emph {et~al.}(2023)\citenamefont {Baggioli}, \citenamefont {Bu},\ and\ \citenamefont {Ziogas}}]{Baggioli:2023tlc}%
  \BibitemOpen
  \bibfield  {author} {\bibinfo {author} {\bibfnamefont {M.}~\bibnamefont {Baggioli}}, \bibinfo {author} {\bibfnamefont {Y.}~\bibnamefont {Bu}}, \ and\ \bibinfo {author} {\bibfnamefont {V.}~\bibnamefont {Ziogas}},\ }\href {\doibase 10.1007/JHEP09(2023)019} {\bibfield  {journal} {\bibinfo  {journal} {JHEP}\ }\textbf {\bibinfo {volume} {09}},\ \bibinfo {pages} {019} (\bibinfo {year} {2023})},\ \Eprint {http://arxiv.org/abs/2304.14173} {arXiv:2304.14173 [hep-th]} \BibitemShut {NoStop}%
\bibitem [{\citenamefont {Liu}\ \emph {et~al.}(2025)\citenamefont {Liu}, \citenamefont {Sun},\ and\ \citenamefont {Wu}}]{Liu:2024tqe}%
  \BibitemOpen
  \bibfield  {author} {\bibinfo {author} {\bibfnamefont {Y.}~\bibnamefont {Liu}}, \bibinfo {author} {\bibfnamefont {Y.-W.}\ \bibnamefont {Sun}}, \ and\ \bibinfo {author} {\bibfnamefont {X.-M.}\ \bibnamefont {Wu}},\ }\href {\doibase 10.1016/j.physc.2025.1354701} {\bibfield  {journal} {\bibinfo  {journal} {Physica C}\ }\textbf {\bibinfo {volume} {632}},\ \bibinfo {pages} {1354701} (\bibinfo {year} {2025})},\ \Eprint {http://arxiv.org/abs/2411.16306} {arXiv:2411.16306 [hep-th]} \BibitemShut {NoStop}%
\bibitem [{\citenamefont {Brattan}\ \emph {et~al.}(2024)\citenamefont {Brattan}, \citenamefont {Matsumoto}, \citenamefont {Baggioli},\ and\ \citenamefont {Amoretti}}]{Brattan:2024dfv}%
  \BibitemOpen
  \bibfield  {author} {\bibinfo {author} {\bibfnamefont {D.~K.}\ \bibnamefont {Brattan}}, \bibinfo {author} {\bibfnamefont {M.}~\bibnamefont {Matsumoto}}, \bibinfo {author} {\bibfnamefont {M.}~\bibnamefont {Baggioli}}, \ and\ \bibinfo {author} {\bibfnamefont {A.}~\bibnamefont {Amoretti}},\ }\href {\doibase 10.1103/PhysRevResearch.6.043097} {\bibfield  {journal} {\bibinfo  {journal} {Phys. Rev. Res.}\ }\textbf {\bibinfo {volume} {6}},\ \bibinfo {pages} {043097} (\bibinfo {year} {2024})},\ \Eprint {http://arxiv.org/abs/2404.05568} {arXiv:2404.05568 [cond-mat.stat-mech]} \BibitemShut {NoStop}%
\bibitem [{\citenamefont {Chen}\ and\ \citenamefont {Lucas}(2017)}]{Chen:2017dsy}%
  \BibitemOpen
  \bibfield  {author} {\bibinfo {author} {\bibfnamefont {C.-F.}\ \bibnamefont {Chen}}\ and\ \bibinfo {author} {\bibfnamefont {A.}~\bibnamefont {Lucas}},\ }\href {\doibase 10.1016/j.physletb.2017.10.023} {\bibfield  {journal} {\bibinfo  {journal} {Phys. Lett. B}\ }\textbf {\bibinfo {volume} {774}},\ \bibinfo {pages} {569} (\bibinfo {year} {2017})},\ \Eprint {http://arxiv.org/abs/1709.01520} {arXiv:1709.01520 [hep-th]} \BibitemShut {NoStop}%
\bibitem [{\citenamefont {Liu}\ and\ \citenamefont {Glorioso}(2018)}]{Liu:2018kfw}%
  \BibitemOpen
  \bibfield  {author} {\bibinfo {author} {\bibfnamefont {H.}~\bibnamefont {Liu}}\ and\ \bibinfo {author} {\bibfnamefont {P.}~\bibnamefont {Glorioso}},\ }\href {\doibase 10.22323/1.305.0008} {\bibfield  {journal} {\bibinfo  {journal} {PoS}\ }\textbf {\bibinfo {volume} {TASI2017}},\ \bibinfo {pages} {008} (\bibinfo {year} {2018})},\ \Eprint {http://arxiv.org/abs/1805.09331} {arXiv:1805.09331 [hep-th]} \BibitemShut {NoStop}%
\bibitem [{\citenamefont {Ammon}\ and\ \citenamefont {Erdmenger}(2015)}]{Ammon_Erdmenger_2015}%
  \BibitemOpen
  \bibfield  {author} {\bibinfo {author} {\bibfnamefont {M.}~\bibnamefont {Ammon}}\ and\ \bibinfo {author} {\bibfnamefont {J.}~\bibnamefont {Erdmenger}},\ }\href@noop {} {\emph {\bibinfo {title} {Gauge/Gravity Duality: Foundations and Applications}}}\ (\bibinfo  {publisher} {Cambridge University Press},\ \bibinfo {year} {2015})\BibitemShut {NoStop}%
\bibitem [{\citenamefont {Zaanen}\ \emph {et~al.}(2015)\citenamefont {Zaanen}, \citenamefont {Liu}, \citenamefont {Sun},\ and\ \citenamefont {Schalm}}]{Zaanen_Liu_Sun_Schalm_2015}%
  \BibitemOpen
  \bibfield  {author} {\bibinfo {author} {\bibfnamefont {J.}~\bibnamefont {Zaanen}}, \bibinfo {author} {\bibfnamefont {Y.}~\bibnamefont {Liu}}, \bibinfo {author} {\bibfnamefont {Y.-W.}\ \bibnamefont {Sun}}, \ and\ \bibinfo {author} {\bibfnamefont {K.}~\bibnamefont {Schalm}},\ }\href@noop {} {\emph {\bibinfo {title} {Holographic Duality in Condensed Matter Physics}}}\ (\bibinfo  {publisher} {Cambridge University Press},\ \bibinfo {year} {2015})\BibitemShut {NoStop}%
\bibitem [{\citenamefont {Natsuume}(2015)}]{Natsuume:2014sfa}%
  \BibitemOpen
  \bibfield  {author} {\bibinfo {author} {\bibfnamefont {M.}~\bibnamefont {Natsuume}},\ }\href {\doibase 10.1007/978-4-431-55441-7} {\emph {\bibinfo {title} {{AdS/CFT Duality User Guide}}}},\ Vol.\ \bibinfo {volume} {903}\ (\bibinfo {year} {2015})\ \Eprint {http://arxiv.org/abs/1409.3575} {arXiv:1409.3575 [hep-th]} \BibitemShut {NoStop}%
\bibitem [{\citenamefont {Baggioli}(2019)}]{Baggioli:2019rrs}%
  \BibitemOpen
  \bibfield  {author} {\bibinfo {author} {\bibfnamefont {M.}~\bibnamefont {Baggioli}},\ }\emph {\bibinfo {title} {{Applied Holography}: {A Practical Mini-Course}}},\ \href {\doibase 10.1007/978-3-030-35184-7} {\bibinfo {type} {Other thesis}},\ \bibinfo  {school} {Madrid, IFT} (\bibinfo {year} {2019}),\ \Eprint {http://arxiv.org/abs/1908.02667} {arXiv:1908.02667 [hep-th]} \BibitemShut {NoStop}%
\bibitem [{\citenamefont {Hartnoll}\ \emph {et~al.}(2016)\citenamefont {Hartnoll}, \citenamefont {Lucas},\ and\ \citenamefont {Sachdev}}]{Hartnoll:2016apf}%
  \BibitemOpen
  \bibfield  {author} {\bibinfo {author} {\bibfnamefont {S.~A.}\ \bibnamefont {Hartnoll}}, \bibinfo {author} {\bibfnamefont {A.}~\bibnamefont {Lucas}}, \ and\ \bibinfo {author} {\bibfnamefont {S.}~\bibnamefont {Sachdev}},\ }\href@noop {} {\  (\bibinfo {year} {2016})},\ \Eprint {http://arxiv.org/abs/1612.07324} {arXiv:1612.07324 [hep-th]} \BibitemShut {NoStop}%
\bibitem [{\citenamefont {Hartnoll}(2009)}]{Hartnoll:2009sz}%
  \BibitemOpen
  \bibfield  {author} {\bibinfo {author} {\bibfnamefont {S.~A.}\ \bibnamefont {Hartnoll}},\ }\href {\doibase 10.1088/0264-9381/26/22/224002} {\bibfield  {journal} {\bibinfo  {journal} {Class. Quant. Grav.}\ }\textbf {\bibinfo {volume} {26}},\ \bibinfo {pages} {224002} (\bibinfo {year} {2009})},\ \Eprint {http://arxiv.org/abs/0903.3246} {arXiv:0903.3246 [hep-th]} \BibitemShut {NoStop}%
\bibitem [{\citenamefont {Donos}\ and\ \citenamefont {Gauntlett}(2015)}]{Donos:2015gia}%
  \BibitemOpen
  \bibfield  {author} {\bibinfo {author} {\bibfnamefont {A.}~\bibnamefont {Donos}}\ and\ \bibinfo {author} {\bibfnamefont {J.~P.}\ \bibnamefont {Gauntlett}},\ }\href {\doibase 10.1103/PhysRevD.92.121901} {\bibfield  {journal} {\bibinfo  {journal} {Phys. Rev. D}\ }\textbf {\bibinfo {volume} {92}},\ \bibinfo {pages} {121901} (\bibinfo {year} {2015})},\ \Eprint {http://arxiv.org/abs/1506.01360} {arXiv:1506.01360 [hep-th]} \BibitemShut {NoStop}%
\bibitem [{\citenamefont {Glorioso}\ \emph {et~al.}(2018)\citenamefont {Glorioso}, \citenamefont {Crossley},\ and\ \citenamefont {Liu}}]{Glorioso:2018mmw}%
  \BibitemOpen
  \bibfield  {author} {\bibinfo {author} {\bibfnamefont {P.}~\bibnamefont {Glorioso}}, \bibinfo {author} {\bibfnamefont {M.}~\bibnamefont {Crossley}}, \ and\ \bibinfo {author} {\bibfnamefont {H.}~\bibnamefont {Liu}},\ }\href@noop {} {\  (\bibinfo {year} {2018})},\ \Eprint {http://arxiv.org/abs/1812.08785} {arXiv:1812.08785 [hep-th]} \BibitemShut {NoStop}%
\bibitem [{\citenamefont {Crossley}\ \emph {et~al.}(2016)\citenamefont {Crossley}, \citenamefont {Glorioso}, \citenamefont {Liu},\ and\ \citenamefont {Wang}}]{Crossley:2015tka}%
  \BibitemOpen
  \bibfield  {author} {\bibinfo {author} {\bibfnamefont {M.}~\bibnamefont {Crossley}}, \bibinfo {author} {\bibfnamefont {P.}~\bibnamefont {Glorioso}}, \bibinfo {author} {\bibfnamefont {H.}~\bibnamefont {Liu}}, \ and\ \bibinfo {author} {\bibfnamefont {Y.}~\bibnamefont {Wang}},\ }\href {\doibase 10.1007/JHEP02(2016)124} {\bibfield  {journal} {\bibinfo  {journal} {JHEP}\ }\textbf {\bibinfo {volume} {02}},\ \bibinfo {pages} {124} (\bibinfo {year} {2016})},\ \Eprint {http://arxiv.org/abs/1504.07611} {arXiv:1504.07611 [hep-th]} \BibitemShut {NoStop}%
\bibitem [{\citenamefont {Haehl}\ \emph {et~al.}(2016)\citenamefont {Haehl}, \citenamefont {Loganayagam},\ and\ \citenamefont {Rangamani}}]{haehl2016fluid}%
  \BibitemOpen
  \bibfield  {author} {\bibinfo {author} {\bibfnamefont {F.~M.}\ \bibnamefont {Haehl}}, \bibinfo {author} {\bibfnamefont {R.}~\bibnamefont {Loganayagam}}, \ and\ \bibinfo {author} {\bibfnamefont {M.}~\bibnamefont {Rangamani}},\ }\href {\doibase 10.1007/JHEP01(2016)184} {\bibfield  {journal} {\bibinfo  {journal} {JHEP}\ }\textbf {\bibinfo {volume} {01}},\ \bibinfo {pages} {184} (\bibinfo {year} {2016})},\ \Eprint {http://arxiv.org/abs/1510.02494} {arXiv:1510.02494 [hep-th]} \BibitemShut {NoStop}%
\bibitem [{\citenamefont {Sakai}\ and\ \citenamefont {Sugimoto}(2005)}]{Sakai:2004cn}%
  \BibitemOpen
  \bibfield  {author} {\bibinfo {author} {\bibfnamefont {T.}~\bibnamefont {Sakai}}\ and\ \bibinfo {author} {\bibfnamefont {S.}~\bibnamefont {Sugimoto}},\ }\href {\doibase 10.1143/PTP.113.843} {\bibfield  {journal} {\bibinfo  {journal} {Prog. Theor. Phys.}\ }\textbf {\bibinfo {volume} {113}},\ \bibinfo {pages} {843} (\bibinfo {year} {2005})},\ \Eprint {http://arxiv.org/abs/hep-th/0412141} {arXiv:hep-th/0412141} \BibitemShut {NoStop}%
\bibitem [{\citenamefont {Varnavides}\ \emph {et~al.}(2023)\citenamefont {Varnavides}, \citenamefont {Yacoby}, \citenamefont {Felser},\ and\ \citenamefont {Narang}}]{Varnavides2023}%
  \BibitemOpen
  \bibfield  {author} {\bibinfo {author} {\bibfnamefont {G.}~\bibnamefont {Varnavides}}, \bibinfo {author} {\bibfnamefont {A.}~\bibnamefont {Yacoby}}, \bibinfo {author} {\bibfnamefont {C.}~\bibnamefont {Felser}}, \ and\ \bibinfo {author} {\bibfnamefont {P.}~\bibnamefont {Narang}},\ }\href {\doibase 10.1038/s41578-023-00597-3} {\bibfield  {journal} {\bibinfo  {journal} {Nature Reviews Materials}\ }\textbf {\bibinfo {volume} {8}},\ \bibinfo {pages} {726} (\bibinfo {year} {2023})}\BibitemShut {NoStop}%
\bibitem [{\citenamefont {Abbasi}\ \emph {et~al.}(2024)\citenamefont {Abbasi}, \citenamefont {Kaminski},\ and\ \citenamefont {Tavakol}}]{PhysRevLett.132.131602}%
  \BibitemOpen
  \bibfield  {author} {\bibinfo {author} {\bibfnamefont {N.}~\bibnamefont {Abbasi}}, \bibinfo {author} {\bibfnamefont {M.}~\bibnamefont {Kaminski}}, \ and\ \bibinfo {author} {\bibfnamefont {O.}~\bibnamefont {Tavakol}},\ }\href {\doibase 10.1103/PhysRevLett.132.131602} {\bibfield  {journal} {\bibinfo  {journal} {Phys. Rev. Lett.}\ }\textbf {\bibinfo {volume} {132}},\ \bibinfo {pages} {131602} (\bibinfo {year} {2024})}\BibitemShut {NoStop}%
\end{thebibliography}
\end{document}